\documentclass[aps,prb,reprint,superscriptaddress,longbibliography, floatfix]{revtex4-2}
\usepackage[T1]{fontenc}
\usepackage{lmodern}
\usepackage{amsmath,amssymb}
\usepackage{graphicx}
\usepackage{hyperref}
\hypersetup{hidelinks,pdfauthor={Ahmed Samir Lotfy et al.},pdftitle={Spectral fingerprints of resonant defect scattering by substitutional Mn in graphene}}

\begin{document}
\title{Spectral fingerprints of resonant defect scattering by substitutional Mn in graphene}
\author{Ahmed Samir Lotfy}
\affiliation{Quantum Solid-State Physics, KU Leuven, 3001 Leuven, Belgium}
\author{Zviadi Zarkua}
\affiliation{Quantum Solid-State Physics, KU Leuven, 3001 Leuven, Belgium}
\author{Renan Villarreal}
\affiliation{Quantum Solid-State Physics, KU Leuven, 3001 Leuven, Belgium}
\author{Rikkie Joris}
\affiliation{Quantum Solid-State Physics, KU Leuven, 3001 Leuven, Belgium}
\author{Muhammad Saad}
\affiliation{Quantum Solid-State Physics, KU Leuven, 3001 Leuven, Belgium}
\author{Koen van Stiphout}
\affiliation{Quantum Solid-State Physics, KU Leuven, 3001 Leuven, Belgium}
\author{Karina Landivar Zambrana}
\affiliation{ETSF and Dipartimento di Fisica ``Aldo Pontremoli'', Universit\`a degli Studi di Milano, Via Celoria 16, I-20133 Milano, Italy}
\author{Steven Brems}
\affiliation{imec vzw (Interuniversitair Micro-Electronica Centrum), 3001 Leuven, Belgium}
\author{Giovanni Di Santo}
\affiliation{Elettra Sincrotrone Trieste, Strada Statale 14 km 163.5, 34149 Trieste, Italy}
\author{Aleksandr Seliverstov}
\affiliation{Quantum Solid-State Physics, KU Leuven, 3001 Leuven, Belgium}
\author{Simona Achilli}
\affiliation{ETSF and Dipartimento di Fisica ``Aldo Pontremoli'', Universit\`a degli Studi di Milano, Via Celoria 16, I-20133 Milano, Italy}
\affiliation{INFN Sezione di Milano and ``European Theoretical Spectroscopy Facility'' (ETSF), Milano, Italy}
\author{Luca Petaccia}
\affiliation{Elettra Sincrotrone Trieste, Strada Statale 14 km 163.5, 34149 Trieste, Italy}
\author{Lino M. C. Pereira}
\email{lino.pereira@kuleuven.be}
\affiliation{Quantum Solid-State Physics, KU Leuven, 3001 Leuven, Belgium}

\date{\today}

\begin{abstract}
Point defects in graphene are predicted to produce different signatures in the momentum-resolved spectral function depending on whether they act as nonresonant, resonant, or sublattice-symmetry-breaking scatterers. Here we test these predictions using substitutional Mn atoms in graphene on Cu(111) as model point scatterers. The substitutional Mn concentration is determined directly by scanning tunneling microscopy, while the corresponding modification of the graphene $\pi$ bands is measured by angle-resolved photoemission spectroscopy. Increasing the Mn concentration from $n_{\mathrm{Mn}}=0$ to $0.177\pm0.013\%$ and $0.44\pm0.03\%$ produces a systematic stretching of the Dirac point region, with a 10--90\% width increasing from $0.24\pm0.03$ to $0.33\pm0.04$ and $0.54\pm0.09$~eV, respectively. Away from the stretched Dirac point region, the momentum linewidth shows no increase with binding energy, whereas its average increases with increasing $n_{\mathrm{Mn}}$, yielding an effective two-dimensional scattering cross section of approximately $5$~nm. Together, the concentration dependence of the Dirac point stretching and the energy independence of the linewidth (away from the Dirac point) are consistent with substitutional Mn acting as a resonant scatterer. By relating changes in quasiparticle dispersion to a well-characterized defect with independently determined concentration, these results establish a quantitative approach to testing theories of resonant impurity scattering in Dirac materials.

\end{abstract}
\maketitle

\section{Introduction}

In ideal monolayer graphene, the $\pi$ bands disperse linearly around the inequivalent $K$ and $K'$ points and meet at the Dirac point energy, where the density of states vanishes. Point defects such as vacancies, adatoms and substitutional impurities perturb these low-energy states and can substantially modify the electronic structure near the Dirac point. Instead of a perfect Dirac cone, defects introduce scattering and localized states that can smear or modify the bands. This can manifest as an elongated or blurred Dirac point region: the two branches appear pulled apart over a finite energy interval, while residual spectral weight persists at the Dirac energy rather than a complete gap opening~\cite{kot2020band}. Such anomalous Dirac point spectra have been widely observed and debated, with interpretations involving lattice imperfections, defect-induced states, many-body interactions, and the substrate~\cite{bostwick2007quasiparticle,zhou2007substrate,zhou2008departure,zhou2008origin,rotenberg2008origin,walter2011effective,avila2013exploring,vita2014understanding,basov2014colloquium,joucken2015charge}. Point defects also increase quasiparticle scattering and broaden the graphene bands. The dispersion and linewidth therefore provide complementary information about the nature of disorder.

Angle-resolved photoemission spectroscopy (ARPES) provides direct insight into these changes in graphene's band structure. Numerous studies on defective graphene or graphene supported on imperfect substrates have reported an extended intensity region in place of a sharp Dirac crossing~\cite{bostwick2007quasiparticle,zhou2007substrate,zhou2008departure,zhou2008origin,rotenberg2008origin,bostwick2010interaction,walter2011effective,avila2013exploring,vita2014understanding,basov2014colloquium,joucken2015charge}. However, the interpretation of such spectra is not unique. Substrate hybridization, interface-induced sublattice inequivalence, and rotational averaging over graphene grains can all produce features resembling a broadened or elongated Dirac point. An apparent elongation by itself is therefore not sufficient to identify resonant disorder. A more stringent test combines the change in dispersion with the associated linewidth, which reflects the imaginary part of the quasiparticle self-energy~\cite{kot2020band,joucken2015charge,bostwick2010interaction}.

The impact of point defects on graphene has been extensively studied using theoretical models. These distinguish resonant defects, nonresonant defects, and defects that break sublattice symmetry on average. Resonant defects create a quasibound state near the Dirac point and a strongly enhanced low-energy scattering amplitude. Early impurity models already proposed that such states could rearrange and smear the Dirac point spectrum~\cite{skrypnyk2007impurity}. More recently, Kot \textit{et al.} established a relation between the defect concentration and the resulting spectral signatures, specifically the stretching of the Dirac point region, $\Delta E_{\mathrm{DP}}$, and the momentum broadening, $\Delta k$~\cite{kot2020band}. The distinction between resonant and nonresonant defects can be expressed in terms of the impurity $T$ matrix, which describes repeated scattering by a single defect potential $V$ in the host:
\begin{equation}
T(E)=V+VG_0(E)T(E),
\label{eq:impurity_T_matrix}
\end{equation}
where $G_0(E)$ is the retarded Green function of the host without the defect. In the low-energy models considered by Kot \textit{et al.}, resonant defects have a singular $T$ matrix near the Dirac point, whereas nonresonant defects have a regular low-energy $T$ matrix~\cite{kot2020band}. This single defect classification is reflected in the concentration dependent spectral response. In their resonant defect examples, $\Delta E_{\mathrm{DP}}$ grows with concentration, while the broadening away from the stretched Dirac point region is relatively uniform in energy. Nonresonant defects instead produce little stretching and a pronounced increase of the linewidth as the energy moves away from the Dirac point. Finally, sublattice-symmetry breaking gives a third, different response, with a true gap and an impurity band. These predictions can be tested by measuring how the spectra change with defect concentration. Indications of such effects can be found in earlier experiments, notably for K adatoms~\cite{bostwick2010interaction} and substitutional N impurities~\cite{joucken2015charge}. For K adatoms, a linear relation was observed between the induced carrier density and the MDC linewidth, both inferred from ARPES. The adatom concentration itself was not determined independently, and Dirac point stretching was not addressed. For substitutional N, the defect concentration was measured directly by scanning tunneling microscopy (STM), and both stretching and broadening were visible, but their dependence on concentration was not analyzed in detail. This motivates a direct experimental comparison of both spectral signatures against an independently measured density of a specific atomic defect.

Here we address these relationships using substitutional Mn in graphene/Cu(111) as a model resonant defect. Previous work established its atomic configuration and STM fingerprint, as well as a preparation approach that retains substitutional Mn while removing most implantation-induced disorder~\cite{lin2021doping,lin2022thermal,villarreal2024achieving}. Our density functional theory (DFT) calculations show that substitutional Mn introduces pronounced impurity-related spectral weight in the Dirac point region, hybridized with the graphene $\pi$ states. The Mn concentration $n_{\mathrm{Mn}}$ is determined independently by STM, while ARPES measures the corresponding $\Delta E_{\mathrm{DP}}$ and $\Delta k$. We observe the combined behavior expected for resonant scattering: the Dirac point stretching increases with $n_{\mathrm{Mn}}$, accompanied by an increased momentum linewidth that is approximately energy independent away from the stretched region. Comparing the defect concentration to the increased momentum linewidth also allows us to obtain an effective single-particle scattering cross section.
\section{Methods}

\subsection{Sample preparation}

The samples consisted of monolayer graphene grown by chemical vapor deposition (CVD) on epitaxial Cu(111) thin films supported on sapphire(0001), following the procedures described previously for Mn-doped graphene on Cu(111) \cite{lin2021doping, lin2022thermal, villarreal2024achieving}. Briefly, epitaxial Cu(111) films were deposited on polished c-plane sapphire substrates, after which graphene was grown by CVD using Ar, H$_2$, and CH$_4$ flows of 5000, 125, and 0.3~sccm, respectively, at a pressure of 750~mbar and at a temperature close to the melting point of Cu. The growth time was 30~min, and both heating and cooling were performed in an Ar/H$_2$ atmosphere.

Three samples were investigated in the present work: one nonimplanted graphene/Cu(111) reference sample and two Mn-implanted samples. Mn implantation was performed using the same ultralow-energy ion implantation approach at room temperature as in Refs. \cite{lin2021doping, lin2022thermal, villarreal2024achieving}. A $^{55}$Mn$^{+}$ ion beam was electrostatically decelerated to a nominal implantation energy of 60~eV and directed onto the sample at normal incidence. The substitutional Mn concentration was varied by changing the implantation fluence while keeping the implantation energy fixed at 60~eV.

After implantation, the samples were annealed in ultrahigh vacuum (UHV) to $700~^{\circ}$C and held at this temperature for 20~min, followed by cooling in UHV without active cooling. All STM and ARPES measurements reported here were performed after this final $700~^{\circ}$C annealing step. This annealing condition was chosen because previous work showed that annealing near $700~^{\circ}$C efficiently removes implantation-induced disorder and surface contamination while retaining substitutional Mn atoms in the graphene lattice, yielding Mn-doped graphene with negligible additional disorder \cite{lin2021doping, lin2022thermal, villarreal2024achieving}.

Exposure to ambient conditions was minimized throughout sample transfer and handling, following the protocol used in Refs. \cite{lin2021doping, lin2022thermal, villarreal2024achieving}. Samples were stored in vacuum and transported under low-vacuum conditions, with air exposure limited to the short times required for mounting and dismounting between experimental setups.

\subsection{Scanning tunneling microscopy}

Scanning tunneling microscopy (STM) measurements were performed in UHV (base pressure $\sim$ $10^{-11}$ mbar) at 78 K with an Omicron low-temperature scanning tunneling microscope (LT-STM) using the same experimental approach as in Refs. \cite{lin2021doping, lin2022thermal, villarreal2024achieving}. All topographies were acquired in constant-current ($I_{\mathrm{tun}}$) mode with a bias voltage applied to the sample ($V_{\mathrm{sample}}$) using electrochemically etched W tips. The oxide layer on the W tips was removed by flash annealing, and the tips were characterized on Au(111) before use.

The concentration of substitutional Mn atoms was determined directly from STM topographies. Substitutional Mn atoms were identified using the characteristic fingerprint established previously for Mn substituting for a carbon atom in graphene on Cu(111): a triangular defect contrast formed by six protrusions associated with neighboring carbon atoms, with the Mn site appearing as a dark region at the center of the structure \cite{lin2021doping, lin2022thermal, villarreal2024achieving}. This assignment was previously validated by comparison between experimental STM topographies and simulated STM images for substitutional Mn and for other possible intrinsic and Mn-related defects \cite{lin2021doping}.

For each implanted sample, substitutional Mn atoms were counted over several atomically resolved STM images. The substitutional concentration was calculated as $n_{\mathrm{Mn}}[\%]=100N_{\mathrm{Mn}}/(A_{\mathrm{STM}}n_{\mathrm{C}})$, where $N_{\mathrm{Mn}}$ is the total number of substitutional Mn atoms counted within the analyzed area $A_{\mathrm{STM}}$, and $n_{\mathrm{C}}=3.92\times10^{15}~\mathrm{cm}^{-2}$ is the areal density of carbon atoms in graphene. Concentrations are reported as percentages of carbon lattice sites. The quoted uncertainties were obtained from the statistical uncertainty of the counting procedure and the standard deviation between different STM images \cite{lin2021doping, lin2022thermal, villarreal2024achieving}.

\subsection{Density functional theory calculations}

DFT calculations were performed within the generalized gradient approximation of Perdew, Burke, and Ernzerhof~\cite{PBE}, using SIESTA with a localized orbital basis and a pseudopotential description of the core electrons~\cite{Sole02}. Graphene was matched to the underlying Cu(111) surface using the theoretical Cu lattice constant, $a_{\mathrm{theo}}=3.59$~\AA, resulting in a 3\% expansion of the overlayer. This commensurate geometry is a computational approximation to the moir\'e-modulated experimental graphene/Cu(111) interface. We considered one substitutional Mn defect in a $(7\times7)$ supercell.%
Periodic replicas normal to the surface were separated by 40~\AA\ of vacuum.

Structural relaxations included three Cu layers, with the two deepest layers kept fixed. Graphene--substrate dispersion interactions were described using the DFT-D2 Grimme correction~\cite{Grimme}. The force tolerance was set to $0.04$~eV/\AA, the real-space mesh cutoff was 450~Ry, and the supercell Brillouin zone was sampled with a $5\times5\times1$ $k$-point mesh for self-consistency and a denser $15\times15\times1$ mesh for the density of states. The calculations were constrained to be non-spin-polarized and provide a nonmagnetic reference electronic structure. One Mn atom per $(7\times7)$ graphene supercell corresponds to approximately 1\% of the carbon sites.

STM simulations were performed within the Tersoff--Hamann approximation~\cite{Tersoff}, assuming a constant tip density of states. The integration window for the electronic density, corresponding to the applied bias, was varied over a 1.0~eV interval around the Fermi energy. Constant-height images were evaluated 3~\AA\ above the graphene plane, with a Gaussian spatial broadening of 1~\AA\ applied to mimic the finite experimental resolution associated with the tip.

\subsection{Angle-resolved photoemission spectroscopy}
\label{sec:arpes-methods}

ARPES measurements were performed at the BaDElPh beamline of Elettra Sincrotrone Trieste, following the acquisition protocol used previously for Mn-doped graphene/Cu(111)~\cite{lin2021doping,lin2022thermal,villarreal2024achieving}. The samples were measured after the final $700\,^{\circ}\mathrm{C}$ anneal, in UHV with a base pressure of approximately $5\times10^{-11}$~mbar. Spectra were acquired at 20~K with a photon energy of $h\nu=34$~eV. The energy and angular resolutions were 10~meV and $\gtrsim0.3^{\circ}$, respectively.

\begin{figure}[tp]
\centering
\includegraphics[width=\linewidth]{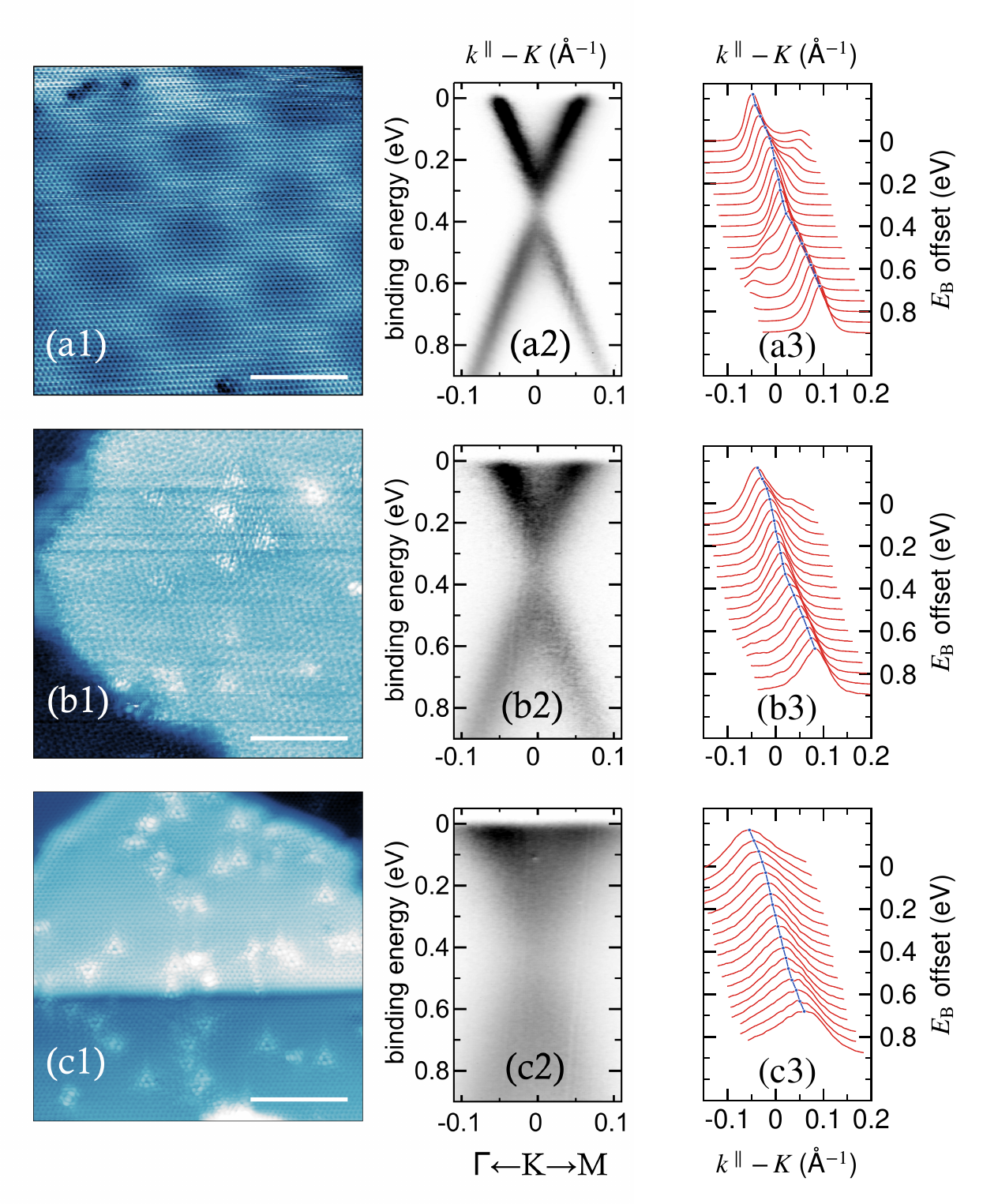}
\caption{STM and ARPES for (a) nonimplanted graphene/Cu(111), (b) $n_{\mathrm{Mn}}=0.177\pm0.013\%$, and (c) $n_{\mathrm{Mn}}=0.44\pm0.03\%$. (a1--c1) STM topographies (scale bar, 5~nm), acquired at $(I_{\mathrm{tun}},V_{\mathrm{sample}})=(0.10~\mathrm{nA},-0.30~\mathrm{V})$, $(0.25~\mathrm{nA},0.15~\mathrm{V})$, and $(105.17~\mathrm{nA},0.23~\mathrm{V})$, respectively. (a2--c2) ARPES overview spectra near $K$ along $\Gamma\!\rightarrow K\!\rightarrow M$, obtained by summing separately acquired $s$- and $p$-polarization data at $h\nu=34$~eV. (a3--c3) Stacked $s$-polarization MDCs from the quantitative analysis. The red curves connect the measured points within the fitting windows; blue symbols mark the fitted Voigt centers. Each curve is normalized to its measured maximum and is offset vertically by its nominal binding energy.}
\label{fig:ARPESSTM}
\end{figure}

Data were acquired separately with linear $p$ and $s$ polarization. Their different intensity distributions arise from photoemission matrix-element effects governed by interference between the two inequivalent carbon sites in the graphene unit cell~\cite{shirley1995brillouin,mucha2008characterization}. Around $K$, the two branches of the Dirac cone have different relative phases on the two sublattices. The transition probability, $I(\mathbf{k},E)\propto|\langle\psi_f|\mathbf{E}\cdot\mathbf{r}|\psi_i\rangle|^2 A(\mathbf{k},E)$, therefore depends strongly on the orientation of the light polarization vector $\mathbf E$ relative to the scattering plane and graphene lattice. The two polarizations emphasize different sides of the same Dirac cone and provide complementary views of the graphene $\pi$ states. The overview spectra in Figs.~\ref{fig:ARPESSTM}(a2--c2) and \ref{fig:ARPESSI}(a) are sums of separately acquired $s$- and $p$-polarization data. Quantitative dispersion and linewidth analysis uses the $s$-polarization MDCs, for which one branch dominates. The additional spectra in Fig.~\ref{fig:ARPESSI}(b,c) use $p$ polarization.

MDCs were extracted at fixed binding energies, using 24 cuts over $0.05$--$1.20$~eV. Each MDC was fitted with one Voigt profile on a linear background, using the native finite data points within $\pm0.12$~\AA$^{-1}$ of the measured intensity maximum. Intensities were normalized to the maximum of the full measured MDC, and equal-weight least squares was used within each momentum window.
The Gaussian FWHM was fixed at $\Delta k_G=0.018$~\AA$^{-1}$ for all samples. The fitted centers $k_{\mathrm{MDC}}$ and Lorentzian FWHM values $\Delta k_L$ were then used to analyze the dispersion and broadening, respectively. The line-shape, resolution, and conditional fit uncertainties are detailed in Appendix~\ref{app:fits}. The mean-linewidth comparison uses the 24 nominal energies from $0.05$ to $1.20$~eV, in steps of $0.05$~eV. 
As in the earlier measurements~\cite{lin2021doping,lin2022thermal,villarreal2024achieving}, spectra were also acquired at the crossing of the graphene $\pi$ band with the Cu $3d$ bands and around the Cu(111) Shockley surface state at $\bar\Gamma$. These additional data and their discussion are included in Appendix~\ref{app:arpes}.
\section{Results and discussion}

\subsection{Determination of the substitutional Mn concentration}

We first use STM to determine the concentration of substitutional Mn defects. Representative topographies of the nonimplanted reference and the two implanted samples after annealing at $700\,^{\circ}\mathrm{C}$ are shown in Fig.~\ref{fig:ARPESSTM}(a1--c1). The nonimplanted graphene/Cu(111) sample shows the expected graphene lattice and graphene/Cu moir\'e modulation. Some dark features, typically attributed to Cu defects, are present~\cite{lin2021doping,lin2022thermal,villarreal2024achieving}. Both implanted samples show triangular defects with the same STM fingerprint [Fig.~\ref{fig:DFT}(a)], reproduced by the simulated image [Fig.~\ref{fig:DFT}(b)] and previously established as resulting from individual Mn atoms substituting for carbon atoms [Fig.~\ref{fig:DFT}(c)]~\cite{lin2021doping,lin2022thermal,villarreal2024achieving}.

Using the counting procedure described above, we determine substitutional Mn concentrations of $n_{\mathrm{Mn}}=0.177\pm0.013\%$ and $0.44\pm0.03\%$ for the lower- and higher-fluence implanted samples, respectively. Expressed as areal densities, these correspond to $(6.9\pm0.5)\times10^{12}$ and $(1.72\pm0.12)\times10^{13}$~cm$^{-2}$. 

The corresponding ARPES maps in Fig.~\ref{fig:ARPESSTM}(a2--c2) show progressive broadening of the graphene $\pi$ bands with increasing Mn concentration. Panels (a3--c3) show the selected MDCs and their fitted centers; these spectral changes are analyzed quantitatively in Sec.~III\,C.

\subsection{Mn density of states near the Dirac point}

\begin{figure}[tbp]
\centering
\includegraphics[width=\columnwidth]{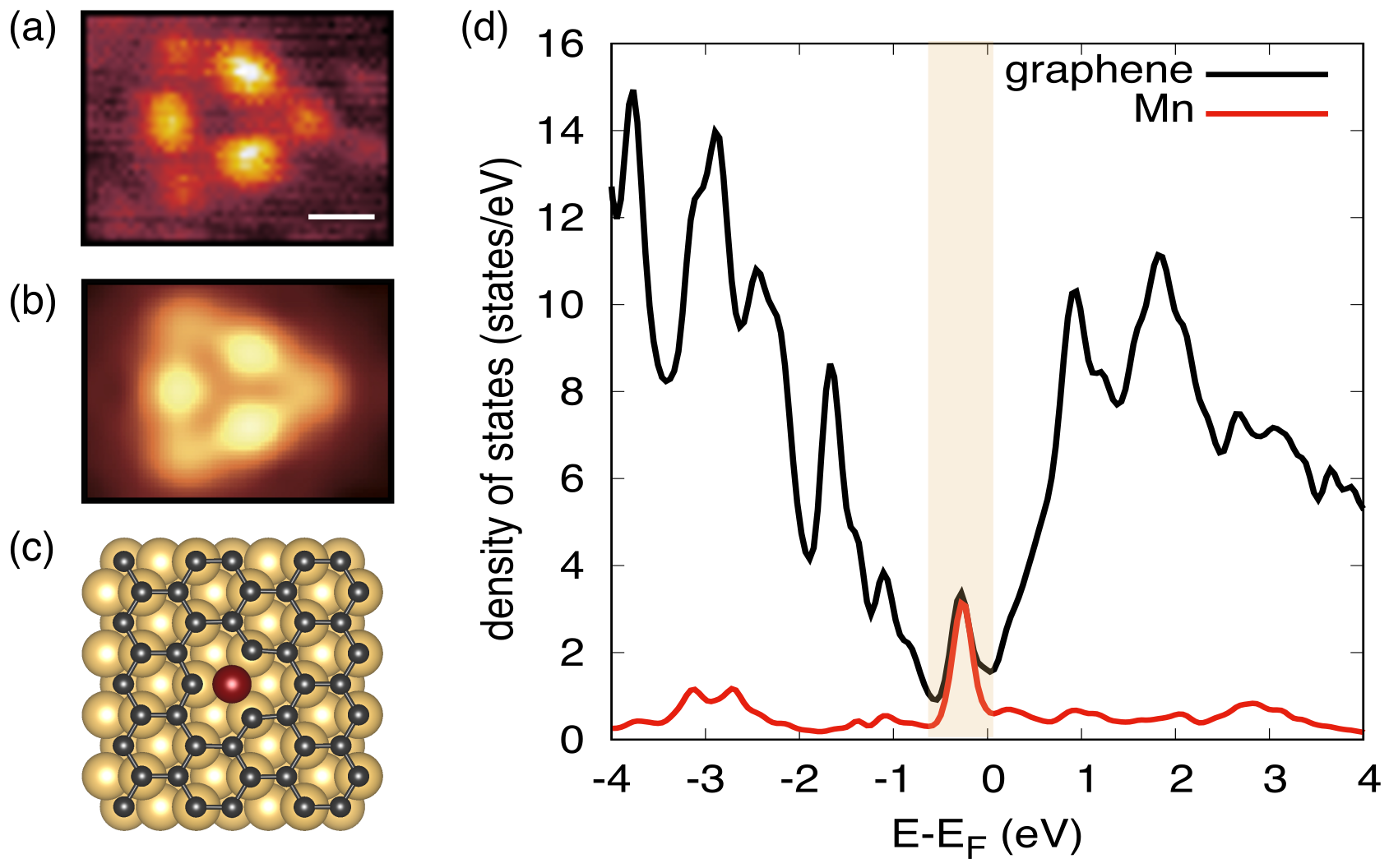}
\caption{Real-space structure and calculated electronic signature of substitutional Mn in graphene/Cu(111). (a) Experimental constant-current STM image of an isolated defect ($I_{\mathrm{tun}}=0.48$~nA, $V_{\mathrm{sample}}=0.42$~V; scale bar, 0.2~nm). (b) DFT-based constant-height STM simulation 3~\AA\ above graphene, reproducing the triangular contrast. (c) Structural model of Mn substituting for a carbon atom. (d) Density of states projected onto the graphene layer (black) and the substitutional Mn atom (red), in states/eV and referenced to $E_{\mathrm F}$. The graphene projection is summed over all carbon atoms in the supercell (not normalized per atom). The Mn-derived peak and finite graphene spectral weight around the Dirac point minimum are consistent with hybridization; the shaded area highlights the relevant energy range.}
\label{fig:DFT}
\end{figure}

Having established the structural assignment of the implanted defects, we next evaluate whether substitutional Mn has the electronic character expected of a strong (resonant) scatterer. Figure~\ref{fig:DFT}(d) shows the density of states projected onto the graphene layer and onto the substitutional Mn atom. The graphene projection retains the characteristic shape associated with the Dirac cone, shifted by approximately $0.3$~eV below $E_{\mathrm F}$ because of charge transfer from Cu(111). At approximately the same energy, the Mn projection exhibits a pronounced impurity-derived peak. The corresponding finite contribution to the graphene-projected density of states indicates that the Mn state hybridizes with the graphene $\pi$ system. The additional spectral weight extends over approximately $0.5$~eV around the Dirac point region.

The calculation therefore predicts that substitutional Mn introduces substantial impurity-derived spectral weight where the density of states of pristine graphene would otherwise be vanishingly small. Coupling between this state and the extended $\pi$ bands can enhance impurity scattering near the Dirac point. Substitutional Mn thus provides a suitable model defect for testing the predicted ARPES fingerprints of resonant scattering. Next, we examine these fingerprints by correlating the independently measured Mn concentration with changes in the dispersion and linewidth.
\subsection{ARPES experiments}

ARPES probes the single-particle spectral function and, through it, the electronic self-energy $\Sigma(\mathbf{k},\omega)$. The real part describes how interactions modify the dispersion relative to the bare band, while the imaginary part describes the loss of quasiparticle coherence through scattering by other excitations, impurities, and defects. Within the quasiparticle framework, the spectral function is~\cite{damascelli2003angle}
\begin{equation}
A(\mathbf{k},\omega)=\frac{1}{\pi}
\frac{|\operatorname{Im}\Sigma(\mathbf{k},\omega)|}
{[\omega-\omega_b(\mathbf{k})-\operatorname{Re}\Sigma(\mathbf{k},\omega)]^2
+[\operatorname{Im}\Sigma(\mathbf{k},\omega)]^2},
\label{eq:spectral_function}
\end{equation}
where $\omega_b(\mathbf{k})$ denotes the bare dispersion. The measured intensity also contains the photoemission matrix element, the Fermi occupation, and experimental broadening. The real part of the self-energy shifts the spectral peak, whereas its imaginary part determines the linewidth and quasiparticle lifetime. For weak momentum dependence of the self-energy and an approximately linear band within the fitted interval, a constant-energy cut is approximately Lorentzian in momentum. With the FWHM convention used here, the Lorentzian momentum width defines an inverse single-particle coherence length, $l^{-1}=\Delta k_L$.

The spectra near the graphene $K$ point show the expected Dirac cone dispersion for all three samples, with progressive broadening as the substitutional Mn concentration increases. The summed $s$- and $p$-polarization data in Fig.~\ref{fig:ARPESSTM}(a2--c2) show both branches of the Dirac cones. For the quantitative analysis we use the $s$-polarization data, for which a single branch dominates [Fig.~\ref{fig:ARPESSTM}(a3--c3)]~\cite{Pramanik2022Anomalies}.%
We apply the same single-Voigt analysis to all three samples to extract the MDC centers and Lorentzian linewidths. The fitted center $k_{\mathrm{MDC}}(E_{\mathrm B})$ follows the dispersion curve, while the Lorentzian width $\Delta k_L$ is related to the momentum broadening.

\subsubsection{Dirac point stretching \texorpdfstring{$\Delta E_{\mathrm{DP}}$}{Delta E (DP)}}

\begin{figure}[tp]
\centering
\includegraphics[width=\columnwidth]{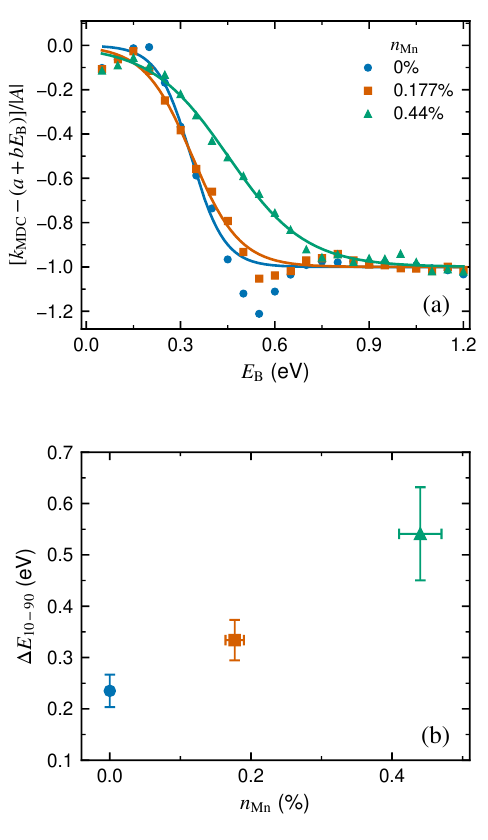}
\caption{Analysis of the Dirac point stretching. (a) Fitted MDC positions after subtraction of the linear dispersion and normalization by the absolute sigmoidal amplitude, $[k-(a+bE_{\mathrm B})]/|A|$. Symbols are extracted positions and solid curves are the fitted sigmoidal terms; concentrations are given in percent. The fits use the full measured range. (b) Stretching $\Delta E_{\mathrm{DP}}\equiv\Delta E_{10-90}=2\ln(9)s$ versus substitutional Mn concentration. Horizontal error bars are the STM concentration uncertainties. Vertical error bars are one-standard-deviation fit uncertainties.}
\label{fig:deltaE}
\end{figure}

To quantify the stretching, the extracted $k_{\mathrm{MDC}}$ values were fitted as a function of binding energy using a linear dispersion plus a sigmoidal deviation,
\begin{equation}
k(E_{\mathrm B})=a+bE_{\mathrm B}+
\frac{A}{1+\exp[-(E_{\mathrm B}-E_0)/s]}.
\label{eq:sigmoid}
\end{equation}
Here $a+bE_{\mathrm B}$ describes the approximately linear overall dispersion, while the sigmoid captures the additional step-like deviation in the Dirac point region. The center $E_0$ and amplitude $A$ characterize the location and magnitude of this deviation. The main quantity of interest is its extent along the binding-energy axis, defined by the 10--90\% transition width,
\begin{equation}
\Delta E_{\mathrm{DP}}\equiv\Delta E_{10-90}=2\ln(9)s.
\label{eq:stretching}
\end{equation}
Weighted nonlinear least squares was used, minimizing the squared residuals divided by the squared MDC-center uncertainties $\delta k_i^2$. The quoted width uncertainties were propagated from the residual-scaled local covariance of $s$, and thus describe the conditional uncertainty within the fitting model; the details are given in Appendix~\ref{app:fits}.

\begin{table*}[t]
\caption{Parameters of the linear-plus-sigmoidal dispersion fits [Eq.~\eqref{eq:sigmoid}]. The uncertainties on $E_0$, $A$, and $\Delta E_{10-90}$ are conditional one-standard-deviation local fit uncertainties, with the covariance scaled by the residual variance. Concentration uncertainties are determined independently by STM.}
\label{tab:sigmoid_fit_parameters}
\begin{ruledtabular}
\begin{tabular}{lccc}
$n_{\mathrm{Mn}}$ (\%) & $E_0$ (eV) & $A$ (\AA$^{-1}$) & $\Delta E_{10-90}$ (eV) \\
\hline
$0$ & $0.333\pm0.008$ & $-0.025\pm0.002$ & $0.24\pm0.03$ \\
$0.177\pm0.013$ & $0.337\pm0.009$ & $-0.030\pm0.003$ & $0.33\pm0.04$ \\
$0.44\pm0.03$ & $0.451\pm0.011$ & $-0.039\pm0.009$ & $0.54\pm0.09$ \\
\end{tabular}
\end{ruledtabular}
\end{table*}

Table~\ref{tab:sigmoid_fit_parameters} summarizes the fit parameters. Figure~\ref{fig:deltaE}(a) compares the extracted positions and fitted curves after subtracting the linear component and normalizing by $|A|$, so that the fitted sigmoid runs from 0 to $-1$. Figure~\ref{fig:deltaE}(b) shows the resulting $\Delta E_{\mathrm{DP}}$ as a function of $n_{\mathrm{Mn}}$. The reference non-implanted sample already exhibits a substantial stretching of $0.24\pm0.03$~eV. This is consistent with the known influence of the graphene--Cu interface and the mesoscale structure of the supported layer. Calculations for graphene on Cu interfaces show that hybridization and local sublattice inequivalence can generate a gaplike separation of order $0.15$--$0.25$~eV~\cite{vita2014understanding}. Nano-ARPES has further shown that individual graphene grains can have a local mini-gap of order 50~meV, while conventional measurements averaging over azimuthally misaligned grains give an apparent separation of order 150~meV~\cite{avila2013exploring}. Related substrate effects have also been observed for graphene on single-crystal Cu~\cite{walter2011copper}. These results show that the observation of Dirac point stretching of a supported sample cannot by itself be taken as evidence of resonant scattering. The stretching dependence on defect concentration is the key aspect that we highlight in the present work. Relative to the nonimplanted reference, $\Delta E_{\mathrm{DP}}$ increases to $0.33\pm0.04$~eV at $n_{\mathrm{Mn}}=0.177\pm0.013\%$ and to $0.54\pm0.09$~eV at $0.44\pm0.03\%$. 
To test whether momentum broadening alone could account for the increased Dirac point stretching, we artificially broaden the reference MDCs until their fitted linewidths match those of the $n_{\mathrm{Mn}}=0.44\pm0.03\%$ sample. Applying the same dispersion analysis to these broadened MDCs gives $\Delta E_{\mathrm{DP}}=0.31$~eV, well below the measured value of $0.54$~eV for the Mn-implanted sample, showing that momentum broadening alone does not account for the full increase in Dirac point stretching (Appendix~\ref{app:broadening-control}).
This growing stretching scale is consistent with the prediction for resonant defects based on the two complementary approaches used in Ref.~\cite{kot2020band}: exact diagonalization of large disordered tight-binding systems and a self-consistent $T$-matrix approximation. For resonant defects such as vacancies or 585 reconstructions, the low-energy $T$ matrix is singular and the Dirac point stretching scales as
$\Delta E_{\mathrm{DP}}\propto\sqrt{n_{\mathrm{def}}/|\ln(cn_{\mathrm{def}})|}$,
where $n_{\mathrm{def}}$ is the defect density and $c$ is related to the microscopic scale. Nonresonant defects, such as the Stone--Wales and paired-vacancy examples considered there, have a nonsingular low-energy $T$ matrix and produce little stretching. The increase observed in Fig.~\ref{fig:deltaE} is therefore perfectly consistent with the resonant defect class. 

The distinction from a sublattice gap is also relevant. In the models considered by Kot \textit{et al.}, preferential occupation of one sublattice produces a true gap accompanied by an impurity band~\cite{kot2020band}. In the present samples, substitutional Mn atoms occupy both graphene sublattices without a detected preference, as established in the earlier STM and implantation studies~\cite{lin2021doping,lin2022thermal,villarreal2024achieving}. The defect distribution therefore preserves sublattice symmetry on average and does not supply a uniform sublattice mass term. The Mn-dependent spectrum is thus described by an elongated and broadened Dirac point region with residual spectral weight. The complementary test of this resonant-scattering interpretation is whether the stretching is accompanied by the expected linewidth behavior, discussed in the next section.
\subsubsection{Momentum broadening and effective cross section}

\begin{figure}[tp]
\centering
\includegraphics[width=\columnwidth]{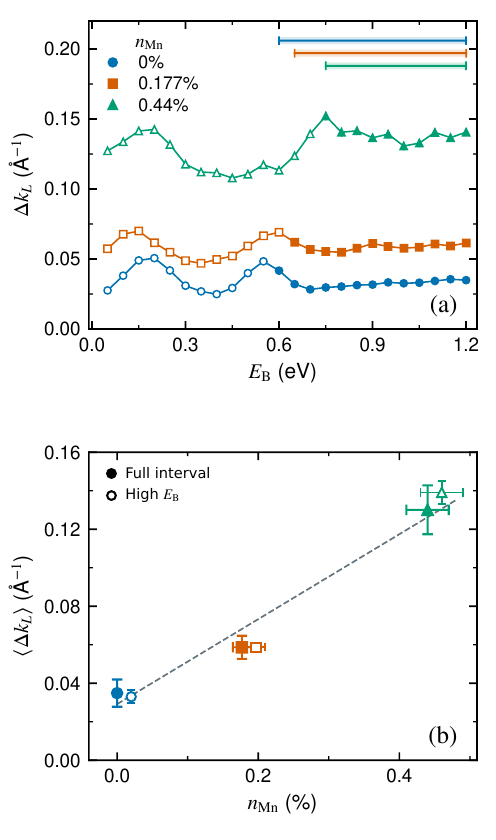}
\caption{Analysis of the momentum broadening. (a) Lorentzian MDC FWHM $\Delta k_L$ versus binding energy for the three Mn concentrations. All points use a Voigt profile on a linear background with Gaussian FWHM $\Delta k_G=0.018$~\AA$^{-1}$ and a momentum half-window of $0.12$~\AA$^{-1}$. Filled symbols and colored bars highlight the higher-binding-energy intervals beyond the stretched Dirac point region, starting at approximately $0.60$, $0.65$, and $0.75$~eV for the reference, lower-, and higher-concentration samples, respectively. These intervals are used for the comparison with the resonant defect linewidth behavior of Ref.~\cite{kot2020band}; open symbols show the remaining measured points. Vertical error bars in (a) are conditional one-standard-deviation local Voigt-fit uncertainties. (b) Mean widths over the full common interval from $0.05$ to $1.20$~eV (filled symbols) and over the highlighted intervals at higher binding energies in (a) (open symbols). For clarity, the open symbols are shifted horizontally by $0.02$ percentage points. Vertical error bars show the sample standard deviation across energy within each averaging interval. Horizontal error bars are the STM concentration uncertainties. The dashed line is the linear fit to the averages over the full interval, with a free intercept, used in Eq.~\eqref{eq:cross-section}.}
\label{fig:deltak}
\end{figure}

The linewidth analysis provides the complementary signature associated with the imaginary part of the self-energy. We use the Lorentzian FWHM $\Delta k_L$ from the same Voigt fits that determine the MDC centers, after accounting for the fixed Gaussian contribution. Representative line shapes and the complete fitting definitions are given in Appendix~\ref{app:fits}. The Gaussian component represents resolution-limited and inhomogeneous contributions, while the Lorentzian component reflects the sample-dependent broadening.

Figure~\ref{fig:deltak}(a) plots the Lorentzian MDC linewidth $\Delta k_L$ as a function of binding energy for the reference and the two Mn-implanted samples. The linewidth is approximately constant except within and around the stretched Dirac point region, where apparent peaks occur near the edges of the upper and lower Dirac cones. At higher binding energies, beyond that region, the linewidth is approximately energy independent: there is no pronounced systematic increase as the energy moves farther from the Dirac point. The filled symbols and colored bars highlight these comparison intervals, starting at about $0.60$, $0.65$, and $0.75$~eV for the reference and the two implanted samples, respectively. The relatively uniform linewidth in these intervals agrees with the resonant defect simulations of Kot \textit{et al.} and contrasts with their simulations for nonresonant defects, which show a strong increase in linewidth as the energy moves away from the Dirac point~\cite{kot2020band}.

Another interesting observation is that the mean linewidth increases strongly with Mn concentration [Fig.~\ref{fig:deltak}(b)]. For this comparison we take the average over the full energy interval, $0.05$--$1.20$~eV. This gives a conservative estimate of the broadening which we can then use to for an effective cross-section analysis.
The means and their standard deviations across energy are listed in Table~\ref{tab:linewidth}.

Figure~\ref{fig:deltak}(b) also shows the linewidths averaged over the highlighted intervals at higher binding energies in Fig.~\ref{fig:deltak}(a). These averages are close to those obtained over the full energy interval and show the same increase with Mn concentration. The main difference is the smaller error bars, since excluding the structured Dirac point region reduces the variation of the linewidth with energy.

\begin{table}[t]
\caption{Mean Lorentzian FWHM and sample standard deviation over the $0.05$--$1.20$~eV interval.}
\label{tab:linewidth}
\begin{ruledtabular}
\begin{tabular}{lcc}
$n_{\mathrm{Mn}}$ (\%) & $\langle\Delta k_L\rangle$ (\AA$^{-1}$) & SD (\AA$^{-1}$) \\
\hline
$0$ & $0.035$ & $0.007$ \\
$0.177\pm0.013$ & $0.059$ & $0.006$ \\
$0.44\pm0.03$ & $0.130$ & $0.013$ \\
\end{tabular}
\end{ruledtabular}
\end{table}
In the dilute-scatterer description, the inverse single-particle coherence length is the product of the impurity areal density and the two-dimensional scattering cross section. We parameterize the concentration dependent widths with an additive model,
\begin{align}
\langle\Delta k_L\rangle&=\Delta k_{L,0}+\rho_{\mathrm{Mn}}\sigma_{\mathrm{eff}},
\label{eq:cross-section}\\
\rho_{\mathrm{Mn}}&=\frac{n_{\mathrm{Mn}}[\%]}{100}\,n_{\mathrm C},
\label{eq:rho}
\end{align}
where the intercept $\Delta k_{L,0}$ represents a fixed, concentration-independent graphene/Cu basline. Its value is fitted together with the slope. The numerical percentage $n_{\mathrm{Mn}}[\%]$ is converted to an areal density using $n_{\mathrm C}=3.92\times10^{15}$~cm$^{-2}=0.392$~\AA$^{-2}$. The cross section consequently has dimensions of length. The linear fit in Fig.~\ref{fig:deltak}(b) has a slope of $0.22$~\AA$^{-1}$ per percentage point
giving an effective single-particle scattering cross section of 
$\sigma_{\mathrm{eff}}=5.6\pm0.9~\mathrm{nm}$ (Appendix~\ref{app:cross-section-error}).
This scale of several nanometers is striking for an atomic substitution. For a weak, short-range perturbation of radius $R$, partial-wave treatments of graphene Dirac fermions give a transport cross section of order $k_FR^2$~\cite{katsnelson2008electron}. Taking $|E_F-E_D|\simeq0.4$~eV, or $k_F\simeq0.6$~nm$^{-1}$, and an atomic-scale radius $R=0.15$--$0.25$~nm gives a scale of only $0.01$--$0.04$~nm. This transport estimate is not quantitatively interchangeable with the ARPES single-particle quantity, but the difference in scale does support the picture of strong scattering rather than a weak geometrical perturbation. Together with the concentration dependent stretching and approximately energy-independent linewidth at higher binding energies beyond the stretched Dirac point region, this supports the assignment of substitutional Mn to the resonant-scattering regime.
The Mn-derived state found in our DFT calculations provides the natural microscopic origin: hybridization with the graphene $\pi$ system creates the local electronic structure that results in enhanced scattering near the Dirac point.

\subsubsection{Broadening in the nonimplanted sample}

The nonimplanted graphene/Cu sample already has a finite linewidth and a sizeable Dirac point stretching. This baseline cannot be straightforwardly attributed to intrinsic atomic point defects in graphene. If the full mean Lorentzian width, $0.035$~\AA$^{-1}$, were due to elastic scattering by point defects, their areal density could be estimated from $\Delta k_L=\rho_{\mathrm{def}}\sigma$. A cross section of 1~nm would require $\rho_{\mathrm{def}}=3.5\times10^{13}$~cm$^{-2}$, or $0.89\%$ of the carbon sites. Even for a much larger cross section of 5~nm, comparable to the Mn-derived effective value, the required density would still be $7.0\times10^{12}$~cm$^{-2}$, or $0.18\%$ of the carbon sites. Atomic vacancies or substitutional defects at these concentrations should be evident in STM, yet no corresponding defect population is observed. This comparison supports the typical interpretation: substrate hybridization, local sublattice inequivalence, rotational averaging over slightly misoriented grains, and other extrinsic or inhomogeneous effects can contribute with such (similar) spectral features~\cite{avila2013exploring,vita2014understanding,walter2011copper}. The observation of an elongated Dirac point alone is therefore insufficient to establish the presence of resonant scattering by point defects. Instead, by introducing and independently counting a controlled population of substitutional Mn, we are able to distinguish its concentration dependent response from this graphene/Cu baseline. The clear fingerprint of resonant scattering is thus the combined characterization of the stretching and broadening effects as a function of defect concentration.
\subsection{Broader implications}

More than simply identifying substitutional Mn as a resonant scatterer, we show how the combination of STM and ARPES allows us to determine how defects with a known atomic structure and concentration change the quasiparticle dispersion and linewidth. Several mechanisms can produce seemingly similar signatures near the Dirac point of supported graphene, making it difficult to interpret the results based only on the shape of the dispersion. We show that correlating the evolution of dispersion and linewidth with an independently measured defect concentration is far more informative. Here, increasing $n_{\mathrm{Mn}}$ produces a larger stretching and additional momentum broadening, with a relatively uniform linewidth at higher binding energies beyond the stretched region, which is the combined behavior predicted for resonant defects.

The independent atomic-scale identification and quantification is central to this comparison. Theoretical descriptions commonly classify vacancies, reconstructed defects, and impurity potentials through their effects on the spectral function and the analytic structure of their $T$ matrices. Experiments often only specify a preparation parameter or measure only some overall measure of disorder. Here, instead, by combining STM-based defect counting with ARPES-based analysis of the dispersion and linewidth, we link a known density of a specific substitutional defect to its spectroscopic response.
This combined approach provides an experimental benchmark for disorder models and can be extended to other substitutional atoms, reconstructed vacancies, or adsorbates to examine the transition between nonresonant and resonant scattering.

\section{Conclusions}

We have used STM and ARPES to determine how substitutional Mn changes the electronic structure of graphene on Cu(111). Identifying and counting the Mn atoms by STM allows us to relate the changes in the graphene $\pi$ bands to the concentration of a specific atomic defect. The DFT calculations show a Mn-derived state near the Dirac point that hybridizes with these bands, providing a microscopic source of enhanced impurity scattering. The Dirac point stretching increases from $0.24\pm0.03$~eV in the nonimplanted reference to $0.33\pm0.04$ and $0.54\pm0.09$~eV at Mn concentrations of $0.177\pm0.013\%$ and $0.44\pm0.03\%$, respectively.
Mn occupies both graphene sublattices without a detectable preference, and the spectra exhibit finite weight through the stretched Dirac point region rather than a real gap. The momentum linewidth also increases with Mn concentration, and beyond the stretched Dirac point region, it remains approximately independent of binding energy. This behavior is consistent with previously reported resonant scattering simulations.
By relating the spectral changes to an independently counted population of substitutional defects, this work establishes a new quantitative approach for testing theories of resonant impurity scattering in Dirac materials.

\begin{acknowledgments}
This work was funded by FWO Vlaanderen and KU Leuven. The authors acknowledge support from the CALIPSOplus project under Grant Agreement No.~730872 from the European Union's Horizon 2020 research and innovation programme. This work also received funding from the European Union's Horizon Europe research and innovation programme under Marie Sk{\l}odowska-Curie Grant Agreement No.~101118915 (TIMES). The authors acknowledge Elettra Sincrotrone Trieste for access to its synchrotron-radiation facilities and thank the staff of the BaDElPh beamline for technical support during the ARPES measurements.
\end{acknowledgments}

\appendix
\section{Line-shape analysis and uncertainty definitions}
\label{app:fits}

\subsection{Voigt fits and resolution contribution}

Each MDC was fitted with a unit-area Voigt profile on a linear background,
\begin{equation}
I(k)=B_0+B_1(k-k_0)+C\,\mathcal V(k-k_0;\sigma_G,\gamma_L),
\label{eq:voigt}
\end{equation}
where $k_0$ is the peak center, $C$ is the integrated peak amplitude, and $\mathcal V$ is the convolution of a Gaussian with standard deviation $\sigma_G$ and a Lorentzian with half width at half maximum $\gamma_L$. The corresponding component FWHM values are
\begin{equation}
\Delta k_G=2\sqrt{2\ln2}\,\sigma_G,
\qquad \Delta k_L=2\gamma_L.
\label{eq:widths}
\end{equation}
The reported $\Delta k_L$ is the Lorentzian component width.

We use a common Gaussian FWHM of $0.018$~\AA$^{-1}$ as the resolution and inhomogeneous-broadening contribution. The nominal angular resolution of $\gtrsim0.3^{\circ}$ corresponds to a momentum width of the order of $0.012$~\AA$^{-1}$ or larger in the present geometry. Additional contributions can originate from angular averaging across the beam spot and small variations in local orientation, strain, or the graphene/Cu environment.

\begin{figure}[tbp]
\centering
\includegraphics[width=\columnwidth]{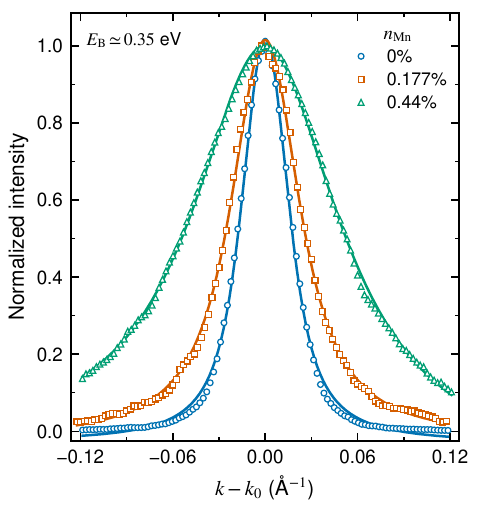}
\caption{Representative $s$-polarization MDCs near $E_{\mathrm B}=0.35$~eV for the three Mn concentrations. Symbols show native measured points and solid lines show the full Voigt-plus-linear-background fits, with $\Delta k_G=0.018$~\AA$^{-1}$ and a half-window of $0.12$~\AA$^{-1}$ about the measured intensity maximum. The momentum axis is centered at the fitted $k_0$, and both measured intensities and fitted curves are divided by the maximum of the corresponding measured MDC.}
\label{fig:voigt}
\end{figure}

Figure~\ref{fig:voigt} illustrates this analysis at $0.35$~eV. The fitted Lorentzian widths are $0.0269\pm0.0004$, $0.0469\pm0.0005$, and $0.1121\pm0.0009$~\AA$^{-1}$ for the reference and the two implanted samples, respectively.

\subsection{Conditional fit uncertainties}

For each MDC, the local parameter covariance was estimated from the least-squares Jacobian $J$ and scaled by the residual variance,
\begin{equation}
\operatorname{Cov}(\mathbf p)=
\frac{\mathrm{RSS}}{N-p}(J^{\mathsf T}J)^{+},
\label{eq:covariance}
\end{equation}
where $N$ is the number of fitted native points, $p$ is the number of free parameters, RSS is the residual sum of squares, and $+$ denotes the pseudoinverse. Diagonal covariance elements supply the squared local uncertainties of $k_0$ and $\Delta k_L$. For the subsequent sigmoid fits, the residuals and Jacobian are weighted by $1/\delta k_i$, and the same residual-variance scaling is applied with the number of energy cuts and five fitted parameters. The stretching uncertainty follows as
\begin{equation}
\delta\Delta E_{10-90}=2\ln(9)\sqrt{\operatorname{Cov}(s,s)}.
\end{equation}

The error bars in Fig.~\ref{fig:deltak}(b) have a different definition. For the $N_E=24$ common energies, the sample mean and sample standard deviation are
\begin{align}
\overline L&=\frac{1}{N_E}\sum_j L_j,\\
s_L^2&=\frac{1}{N_E-1}\sum_j(L_j-\overline L)^2,
\label{eq:energy-spread}
\end{align}
where $L_j=\Delta k_L(E_j)$. This spread describes the variation over energy.%

\subsection{Propagation for the effective cross section}
\label{app:cross-section-error}

The dashed line in Fig.~\ref{fig:deltak}(b) is an unweighted least-squares fit with a free intercept. Writing $x_i=n_{\mathrm{Mn},i}[\%]$ and $y_i=\overline L_i$ for the three samples, its slope is
\begin{equation}
m=\frac{\sum_i(x_i-\bar x)(y_i-\bar y)}{D},
\qquad D=\sum_i(x_i-\bar x)^2.
\label{eq:ols-slope}
\end{equation}
The effective cross section in nanometers is $\sigma_{\mathrm{eff}}=100m/(10n_{\mathrm C})$ with $n_{\mathrm C}=0.392$~\AA$^{-2}$ and $m$ expressed in \AA$^{-1}$ per percentage point.

To obtain a conservative measure of the variation over the whole energy range, we propagate the standard deviation across energy $s_{L,i}$ as a linewidth spread together with the STM concentration uncertainty $\delta x_i$. Assuming independent sample inputs, first-order propagation gives
\begin{equation}
(\delta m)^2=\sum_i\left[
\left(\frac{\partial m}{\partial x_i}\delta x_i\right)^2+
\left(\frac{\partial m}{\partial y_i}s_{L,i}\right)^2\right],
\label{eq:slope-error}
\end{equation}
with
\begin{align}
\frac{\partial m}{\partial y_i}&=\frac{x_i-\bar x}{D},\\
\frac{\partial m}{\partial x_i}&=
\frac{(y_i-\bar y)-2m(x_i-\bar x)}{D}.
\end{align}
This yields $\sigma_{\mathrm{eff}}=5.6\pm0.9$~nm.
\section{Additional ARPES data}
\label{app:arpes}

\begin{figure}[tbp]
\centering
\includegraphics[width=\columnwidth]{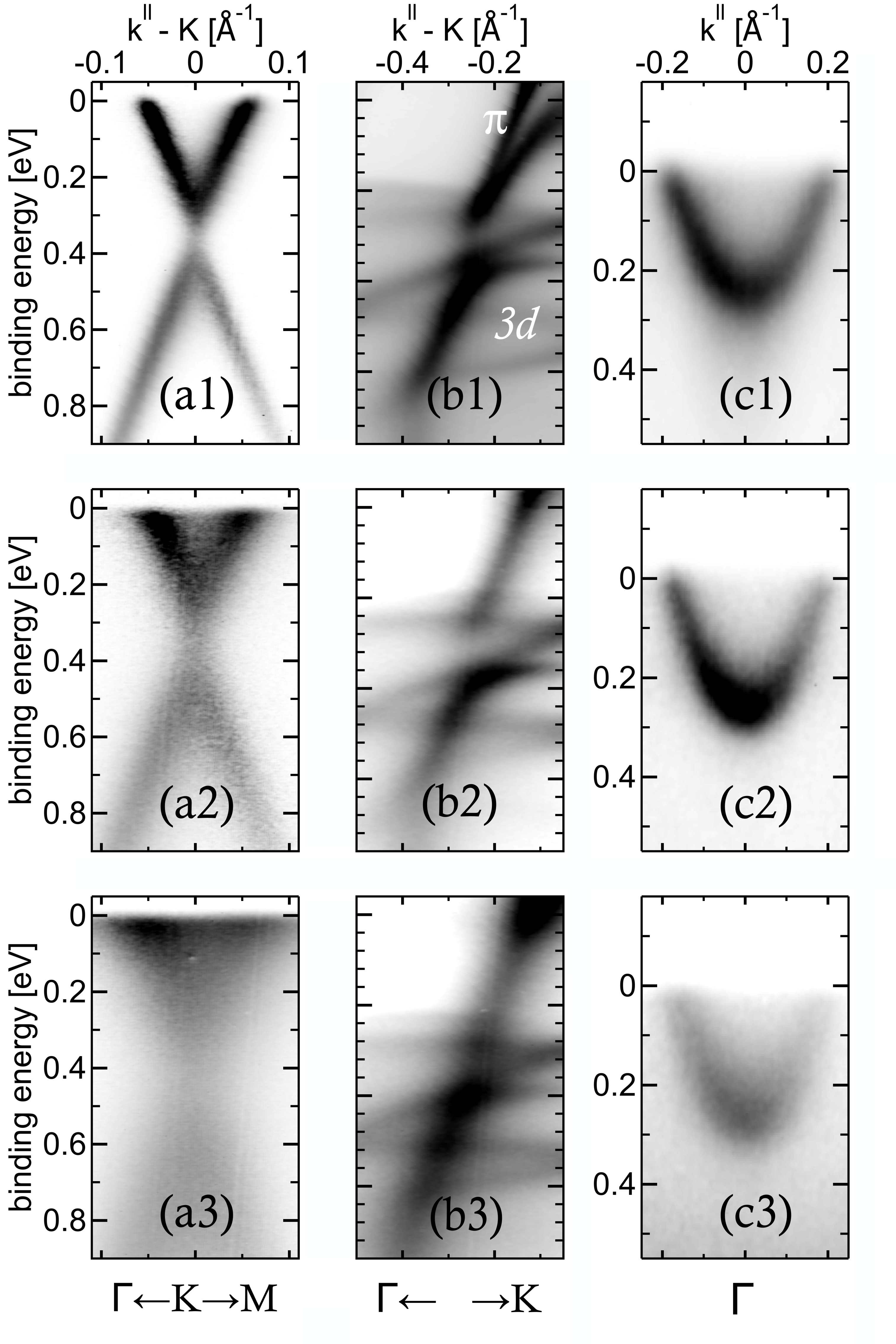}
\caption{Additional ARPES spectra ($h\nu=34$~eV) for (1) nonimplanted graphene/Cu(111), (2) $n_{\mathrm{Mn}}=0.177\pm0.013\%$, and (3) $n_{\mathrm{Mn}}=0.44\pm0.03\%$. (a) Dirac point region along $\Gamma\!\rightarrow K\!\rightarrow M$, shown as the sum of $s$- and $p$-polarization data. (b) Crossing of the graphene $\pi$ band with the Cu $3d$ bands, where the discontinuity is a hybridization-induced avoided crossing. (c) Cu(111) Shockley surface state near $\bar\Gamma$. Panels (b,c) use $p$ polarization.}
\label{fig:ARPESSI}
\end{figure}

In addition to the spectra near the Dirac point, Fig.~\ref{fig:ARPESSI} shows two regions sensitive to the graphene--Cu interface. In Fig.~\ref{fig:ARPESSI}(b), the graphene $\pi$ band crosses the Cu-derived $3d$ bands; the avoided crossing is a signature of their hybridization. Figure~\ref{fig:ARPESSI}(c) shows the Cu(111) Shockley surface state near $\bar\Gamma$.

The nonimplanted and lower-concentration samples behave similarly in both regions. The avoided crossing and the Shockley state remain visible, indicating little detectable change in the Cu-derived bands or overall graphene--Cu hybridization at the lower Mn concentration. At $n_{\mathrm{Mn}}=0.44\pm0.03\%$, the avoided crossing is unresolved or strongly suppressed and the Shockley state is weaker. This loss of spectral definition is consistent with stronger defect-induced scattering and reduced coherence, and may also involve changes in the graphene--Cu hybridization. Related changes in these interface-sensitive features have been observed for irradiation-induced defects in graphene/Cu(111)~\cite{zarkua2026electronic}. 
\section{Momentum-broadening test}
\label{app:broadening-control}

An increase in the MDC linewidth can shift a fitted single-component center when the measured profile contains asymmetric or overlapping spectral weight. To test whether this effect can artificially produce the stretching observed in the implanted samples (i.e. mimic the effect of resonant scattering), we broadened the measured $s$-polarization MDCs of the reference (non-implanted) sample, along momentum, and then applied the dispersion analysis.
At each energy, the reference profile was convolved with a symmetric Lorentzian whose FWHM was adjusted until the refitted Lorentzian linewidth matched that measured for $n_{\mathrm{Mn}}=0.44\pm0.03\%$.
The broadened MDCs were analyzed with the same single-Voigt profile and linear background as the measured data, fixing $\Delta k_G=0.018$~\AA$^{-1}$.
The extracted centers were then fitted with the weighted linear-plus-sigmoidal model of Eq.~\eqref{eq:sigmoid}.
The reference gives $\Delta E_{\mathrm{DP}}=0.24$~eV, increasing to $0.31$~eV after our added momentum broadening, whereas the measured highest-Mn sample gives $0.54$~eV.
Additional momentum broadening therefore does produces some apparent stretching, but that effect is much smaller than the stretching observed in the implanted sample.
\bibliography{references}

\begin{thebibliography}{27}%
\makeatletter
\providecommand \@ifxundefined [1]{%
 \@ifx{#1\undefined}
}%
\providecommand \@ifnum [1]{%
 \ifnum #1\expandafter \@firstoftwo
 \else \expandafter \@secondoftwo
 \fi
}%
\providecommand \@ifx [1]{%
 \ifx #1\expandafter \@firstoftwo
 \else \expandafter \@secondoftwo
 \fi
}%
\providecommand \natexlab [1]{#1}%
\providecommand \enquote  [1]{``#1''}%
\providecommand \bibnamefont  [1]{#1}%
\providecommand \bibfnamefont [1]{#1}%
\providecommand \citenamefont [1]{#1}%
\providecommand \href@noop [0]{\@secondoftwo}%
\providecommand \href [0]{\begingroup \@sanitize@url \@href}%
\providecommand \@href[1]{\@@startlink{#1}\@@href}%
\providecommand \@@href[1]{\endgroup#1\@@endlink}%
\providecommand \@sanitize@url [0]{\catcode `\\12\catcode `\$12\catcode `\&12\catcode `\#12\catcode `\^12\catcode `\_12\catcode `\%12\relax}%
\providecommand \@@startlink[1]{}%
\providecommand \@@endlink[0]{}%
\providecommand \url  [0]{\begingroup\@sanitize@url \@url }%
\providecommand \@url [1]{\endgroup\@href {#1}{\urlprefix }}%
\providecommand \urlprefix  [0]{URL }%
\providecommand \Eprint [0]{\href }%
\providecommand \doibase [0]{https://doi.org/}%
\providecommand \selectlanguage [0]{\@gobble}%
\providecommand \bibinfo  [0]{\@secondoftwo}%
\providecommand \bibfield  [0]{\@secondoftwo}%
\providecommand \translation [1]{[#1]}%
\providecommand \BibitemOpen [0]{}%
\providecommand \bibitemStop [0]{}%
\providecommand \bibitemNoStop [0]{.\EOS\space}%
\providecommand \EOS [0]{\spacefactor3000\relax}%
\providecommand \BibitemShut  [1]{\csname bibitem#1\endcsname}%
\let\auto@bib@innerbib\@empty
\bibitem [{\citenamefont {Kot}\ \emph {et~al.}(2020)\citenamefont {Kot}, \citenamefont {Parnell}, \citenamefont {Habibian}, \citenamefont {Stra{\ss}er}, \citenamefont {Ostrovsky},\ and\ \citenamefont {Ast}}]{kot2020band}%
  \BibitemOpen
  \bibfield  {author} {\bibinfo {author} {\bibfnamefont {P.}~\bibnamefont {Kot}}, \bibinfo {author} {\bibfnamefont {J.}~\bibnamefont {Parnell}}, \bibinfo {author} {\bibfnamefont {S.}~\bibnamefont {Habibian}}, \bibinfo {author} {\bibfnamefont {C.}~\bibnamefont {Stra{\ss}er}}, \bibinfo {author} {\bibfnamefont {P.~M.}\ \bibnamefont {Ostrovsky}},\ and\ \bibinfo {author} {\bibfnamefont {C.~R.}\ \bibnamefont {Ast}},\ }\bibfield  {title} {\bibinfo {title} {Band dispersion of graphene with structural defects},\ }\href {https://doi.org/10.1103/PhysRevB.101.235116} {\bibfield  {journal} {\bibinfo  {journal} {Phys. Rev. B}\ }\textbf {\bibinfo {volume} {101}},\ \bibinfo {pages} {235116} (\bibinfo {year} {2020})}\BibitemShut {NoStop}%
\bibitem [{\citenamefont {Bostwick}\ \emph {et~al.}(2007)\citenamefont {Bostwick}, \citenamefont {Ohta}, \citenamefont {Seyller}, \citenamefont {Horn},\ and\ \citenamefont {Rotenberg}}]{bostwick2007quasiparticle}%
  \BibitemOpen
  \bibfield  {author} {\bibinfo {author} {\bibfnamefont {A.}~\bibnamefont {Bostwick}}, \bibinfo {author} {\bibfnamefont {T.}~\bibnamefont {Ohta}}, \bibinfo {author} {\bibfnamefont {T.}~\bibnamefont {Seyller}}, \bibinfo {author} {\bibfnamefont {K.}~\bibnamefont {Horn}},\ and\ \bibinfo {author} {\bibfnamefont {E.}~\bibnamefont {Rotenberg}},\ }\bibfield  {title} {\bibinfo {title} {Quasiparticle dynamics in graphene},\ }\href {https://doi.org/10.1038/nphys477} {\bibfield  {journal} {\bibinfo  {journal} {Nat. Phys.}\ }\textbf {\bibinfo {volume} {3}},\ \bibinfo {pages} {36} (\bibinfo {year} {2007})}\BibitemShut {NoStop}%
\bibitem [{\citenamefont {Zhou}\ \emph {et~al.}(2007)\citenamefont {Zhou}, \citenamefont {Gweon}, \citenamefont {Fedorov}, \citenamefont {First}, \citenamefont {de~Heer}, \citenamefont {Lee}, \citenamefont {Guinea}, \citenamefont {Castro~Neto},\ and\ \citenamefont {Lanzara}}]{zhou2007substrate}%
  \BibitemOpen
  \bibfield  {author} {\bibinfo {author} {\bibfnamefont {S.~Y.}\ \bibnamefont {Zhou}}, \bibinfo {author} {\bibfnamefont {G.-H.}\ \bibnamefont {Gweon}}, \bibinfo {author} {\bibfnamefont {A.~V.}\ \bibnamefont {Fedorov}}, \bibinfo {author} {\bibfnamefont {P.~N.}\ \bibnamefont {First}}, \bibinfo {author} {\bibfnamefont {W.~A.}\ \bibnamefont {de~Heer}}, \bibinfo {author} {\bibfnamefont {D.-H.}\ \bibnamefont {Lee}}, \bibinfo {author} {\bibfnamefont {F.}~\bibnamefont {Guinea}}, \bibinfo {author} {\bibfnamefont {A.~H.}\ \bibnamefont {Castro~Neto}},\ and\ \bibinfo {author} {\bibfnamefont {A.}~\bibnamefont {Lanzara}},\ }\bibfield  {title} {\bibinfo {title} {Substrate-induced bandgap opening in epitaxial graphene},\ }\href {https://doi.org/10.1038/nmat2003} {\bibfield  {journal} {\bibinfo  {journal} {Nat. Mater.}\ }\textbf {\bibinfo {volume} {6}},\ \bibinfo {pages} {770} (\bibinfo {year} {2007})}\BibitemShut {NoStop}%
\bibitem [{\citenamefont {Zhou}\ \emph {et~al.}(2008{\natexlab{a}})\citenamefont {Zhou}, \citenamefont {Siegel}, \citenamefont {Fedorov},\ and\ \citenamefont {Lanzara}}]{zhou2008departure}%
  \BibitemOpen
  \bibfield  {author} {\bibinfo {author} {\bibfnamefont {S.~Y.}\ \bibnamefont {Zhou}}, \bibinfo {author} {\bibfnamefont {D.~A.}\ \bibnamefont {Siegel}}, \bibinfo {author} {\bibfnamefont {A.~V.}\ \bibnamefont {Fedorov}},\ and\ \bibinfo {author} {\bibfnamefont {A.}~\bibnamefont {Lanzara}},\ }\bibfield  {title} {\bibinfo {title} {Departure from the conical dispersion in epitaxial graphene},\ }\href {https://doi.org/10.1016/j.physe.2007.10.121} {\bibfield  {journal} {\bibinfo  {journal} {Physica E}\ }\textbf {\bibinfo {volume} {40}},\ \bibinfo {pages} {2642} (\bibinfo {year} {2008}{\natexlab{a}})}\BibitemShut {NoStop}%
\bibitem [{\citenamefont {Zhou}\ \emph {et~al.}(2008{\natexlab{b}})\citenamefont {Zhou}, \citenamefont {Siegel}, \citenamefont {Fedorov}, \citenamefont {El~Gabaly}, \citenamefont {Schmid}, \citenamefont {Castro~Neto}, \citenamefont {Lee},\ and\ \citenamefont {Lanzara}}]{zhou2008origin}%
  \BibitemOpen
  \bibfield  {author} {\bibinfo {author} {\bibfnamefont {S.~Y.}\ \bibnamefont {Zhou}}, \bibinfo {author} {\bibfnamefont {D.~A.}\ \bibnamefont {Siegel}}, \bibinfo {author} {\bibfnamefont {A.~V.}\ \bibnamefont {Fedorov}}, \bibinfo {author} {\bibfnamefont {F.}~\bibnamefont {El~Gabaly}}, \bibinfo {author} {\bibfnamefont {A.~K.}\ \bibnamefont {Schmid}}, \bibinfo {author} {\bibfnamefont {A.~H.}\ \bibnamefont {Castro~Neto}}, \bibinfo {author} {\bibfnamefont {D.-H.}\ \bibnamefont {Lee}},\ and\ \bibinfo {author} {\bibfnamefont {A.}~\bibnamefont {Lanzara}},\ }\bibfield  {title} {\bibinfo {title} {Origin of the energy bandgap in epitaxial graphene},\ }\href {https://doi.org/10.1038/nmat2154b} {\bibfield  {journal} {\bibinfo  {journal} {Nat. Mater.}\ }\textbf {\bibinfo {volume} {7}},\ \bibinfo {pages} {259} (\bibinfo {year} {2008}{\natexlab{b}})}\BibitemShut {NoStop}%
\bibitem [{\citenamefont {Rotenberg}\ \emph {et~al.}(2008)\citenamefont {Rotenberg}, \citenamefont {Bostwick}, \citenamefont {Ohta}, \citenamefont {McChesney}, \citenamefont {Seyller},\ and\ \citenamefont {Horn}}]{rotenberg2008origin}%
  \BibitemOpen
  \bibfield  {author} {\bibinfo {author} {\bibfnamefont {E.}~\bibnamefont {Rotenberg}}, \bibinfo {author} {\bibfnamefont {A.}~\bibnamefont {Bostwick}}, \bibinfo {author} {\bibfnamefont {T.}~\bibnamefont {Ohta}}, \bibinfo {author} {\bibfnamefont {J.~L.}\ \bibnamefont {McChesney}}, \bibinfo {author} {\bibfnamefont {T.}~\bibnamefont {Seyller}},\ and\ \bibinfo {author} {\bibfnamefont {K.}~\bibnamefont {Horn}},\ }\bibfield  {title} {\bibinfo {title} {Origin of the energy bandgap in epitaxial graphene},\ }\href {https://doi.org/10.1038/nmat2154a} {\bibfield  {journal} {\bibinfo  {journal} {Nat. Mater.}\ }\textbf {\bibinfo {volume} {7}},\ \bibinfo {pages} {258} (\bibinfo {year} {2008})}\BibitemShut {NoStop}%
\bibitem [{\citenamefont {Walter}\ \emph {et~al.}(2011{\natexlab{a}})\citenamefont {Walter}, \citenamefont {Bostwick}, \citenamefont {Jeon}, \citenamefont {Speck}, \citenamefont {Ostler}, \citenamefont {Seyller}, \citenamefont {Moreschini}, \citenamefont {Chang}, \citenamefont {Polini}, \citenamefont {Asgari}, \citenamefont {MacDonald}, \citenamefont {Horn},\ and\ \citenamefont {Rotenberg}}]{walter2011effective}%
  \BibitemOpen
  \bibfield  {author} {\bibinfo {author} {\bibfnamefont {A.~L.}\ \bibnamefont {Walter}}, \bibinfo {author} {\bibfnamefont {A.}~\bibnamefont {Bostwick}}, \bibinfo {author} {\bibfnamefont {K.-J.}\ \bibnamefont {Jeon}}, \bibinfo {author} {\bibfnamefont {F.}~\bibnamefont {Speck}}, \bibinfo {author} {\bibfnamefont {M.}~\bibnamefont {Ostler}}, \bibinfo {author} {\bibfnamefont {T.}~\bibnamefont {Seyller}}, \bibinfo {author} {\bibfnamefont {L.}~\bibnamefont {Moreschini}}, \bibinfo {author} {\bibfnamefont {Y.~J.}\ \bibnamefont {Chang}}, \bibinfo {author} {\bibfnamefont {M.}~\bibnamefont {Polini}}, \bibinfo {author} {\bibfnamefont {R.}~\bibnamefont {Asgari}}, \bibinfo {author} {\bibfnamefont {A.~H.}\ \bibnamefont {MacDonald}}, \bibinfo {author} {\bibfnamefont {K.}~\bibnamefont {Horn}},\ and\ \bibinfo {author} {\bibfnamefont {E.}~\bibnamefont {Rotenberg}},\ }\bibfield  {title} {\bibinfo {title} {Effective screening and the plasmaron bands in graphene},\ }\href {https://doi.org/10.1103/PhysRevB.84.085410} {\bibfield
  {journal} {\bibinfo  {journal} {Phys. Rev. B}\ }\textbf {\bibinfo {volume} {84}},\ \bibinfo {pages} {085410} (\bibinfo {year} {2011}{\natexlab{a}})}\BibitemShut {NoStop}%
\bibitem [{\citenamefont {Avila}\ \emph {et~al.}(2013)\citenamefont {Avila}, \citenamefont {Razado}, \citenamefont {Lorcy}, \citenamefont {Fleurier}, \citenamefont {Pichonat}, \citenamefont {Vignaud}, \citenamefont {Wallart},\ and\ \citenamefont {Asensio}}]{avila2013exploring}%
  \BibitemOpen
  \bibfield  {author} {\bibinfo {author} {\bibfnamefont {J.}~\bibnamefont {Avila}}, \bibinfo {author} {\bibfnamefont {I.}~\bibnamefont {Razado}}, \bibinfo {author} {\bibfnamefont {S.}~\bibnamefont {Lorcy}}, \bibinfo {author} {\bibfnamefont {R.}~\bibnamefont {Fleurier}}, \bibinfo {author} {\bibfnamefont {E.}~\bibnamefont {Pichonat}}, \bibinfo {author} {\bibfnamefont {D.}~\bibnamefont {Vignaud}}, \bibinfo {author} {\bibfnamefont {X.}~\bibnamefont {Wallart}},\ and\ \bibinfo {author} {\bibfnamefont {M.~C.}\ \bibnamefont {Asensio}},\ }\bibfield  {title} {\bibinfo {title} {Exploring electronic structure of one-atom thick polycrystalline graphene films: A nano angle resolved photoemission study},\ }\href {https://doi.org/10.1038/srep02439} {\bibfield  {journal} {\bibinfo  {journal} {Sci. Rep.}\ }\textbf {\bibinfo {volume} {3}},\ \bibinfo {pages} {2439} (\bibinfo {year} {2013})}\BibitemShut {NoStop}%
\bibitem [{\citenamefont {Vita}\ \emph {et~al.}(2014)\citenamefont {Vita}, \citenamefont {B{\"o}ttcher}, \citenamefont {Horn}, \citenamefont {Voloshina}, \citenamefont {Ovcharenko}, \citenamefont {Kampen}, \citenamefont {Thissen},\ and\ \citenamefont {Dedkov}}]{vita2014understanding}%
  \BibitemOpen
  \bibfield  {author} {\bibinfo {author} {\bibfnamefont {H.}~\bibnamefont {Vita}}, \bibinfo {author} {\bibfnamefont {S.}~\bibnamefont {B{\"o}ttcher}}, \bibinfo {author} {\bibfnamefont {K.}~\bibnamefont {Horn}}, \bibinfo {author} {\bibfnamefont {E.~N.}\ \bibnamefont {Voloshina}}, \bibinfo {author} {\bibfnamefont {R.~E.}\ \bibnamefont {Ovcharenko}}, \bibinfo {author} {\bibfnamefont {T.}~\bibnamefont {Kampen}}, \bibinfo {author} {\bibfnamefont {A.}~\bibnamefont {Thissen}},\ and\ \bibinfo {author} {\bibfnamefont {Y.}~\bibnamefont {Dedkov}},\ }\bibfield  {title} {\bibinfo {title} {Understanding the origin of band gap formation in graphene on metals: Graphene on {Cu/Ir(111)}},\ }\href {https://doi.org/10.1038/srep05704} {\bibfield  {journal} {\bibinfo  {journal} {Sci. Rep.}\ }\textbf {\bibinfo {volume} {4}},\ \bibinfo {pages} {5704} (\bibinfo {year} {2014})}\BibitemShut {NoStop}%
\bibitem [{\citenamefont {Basov}\ \emph {et~al.}(2014)\citenamefont {Basov}, \citenamefont {Fogler}, \citenamefont {Lanzara}, \citenamefont {Wang},\ and\ \citenamefont {Zhang}}]{basov2014colloquium}%
  \BibitemOpen
  \bibfield  {author} {\bibinfo {author} {\bibfnamefont {D.~N.}\ \bibnamefont {Basov}}, \bibinfo {author} {\bibfnamefont {M.~M.}\ \bibnamefont {Fogler}}, \bibinfo {author} {\bibfnamefont {A.}~\bibnamefont {Lanzara}}, \bibinfo {author} {\bibfnamefont {F.}~\bibnamefont {Wang}},\ and\ \bibinfo {author} {\bibfnamefont {Y.}~\bibnamefont {Zhang}},\ }\bibfield  {title} {\bibinfo {title} {Colloquium: Graphene spectroscopy},\ }\href {https://doi.org/10.1103/RevModPhys.86.959} {\bibfield  {journal} {\bibinfo  {journal} {Rev. Mod. Phys.}\ }\textbf {\bibinfo {volume} {86}},\ \bibinfo {pages} {959} (\bibinfo {year} {2014})}\BibitemShut {NoStop}%
\bibitem [{\citenamefont {Joucken}\ \emph {et~al.}(2015)\citenamefont {Joucken}, \citenamefont {Tison}, \citenamefont {Le~F{\`e}vre}, \citenamefont {Tejeda}, \citenamefont {Taleb-Ibrahimi}, \citenamefont {Conrad}, \citenamefont {Repain}, \citenamefont {Chacon}, \citenamefont {Bellec}, \citenamefont {Girard}, \citenamefont {Rousset}, \citenamefont {Ghijsen}, \citenamefont {Sporken}, \citenamefont {Amara}, \citenamefont {Ducastelle},\ and\ \citenamefont {Lagoute}}]{joucken2015charge}%
  \BibitemOpen
  \bibfield  {author} {\bibinfo {author} {\bibfnamefont {F.}~\bibnamefont {Joucken}}, \bibinfo {author} {\bibfnamefont {Y.}~\bibnamefont {Tison}}, \bibinfo {author} {\bibfnamefont {P.}~\bibnamefont {Le~F{\`e}vre}}, \bibinfo {author} {\bibfnamefont {A.}~\bibnamefont {Tejeda}}, \bibinfo {author} {\bibfnamefont {A.}~\bibnamefont {Taleb-Ibrahimi}}, \bibinfo {author} {\bibfnamefont {E.}~\bibnamefont {Conrad}}, \bibinfo {author} {\bibfnamefont {V.}~\bibnamefont {Repain}}, \bibinfo {author} {\bibfnamefont {C.}~\bibnamefont {Chacon}}, \bibinfo {author} {\bibfnamefont {A.}~\bibnamefont {Bellec}}, \bibinfo {author} {\bibfnamefont {Y.}~\bibnamefont {Girard}}, \bibinfo {author} {\bibfnamefont {S.}~\bibnamefont {Rousset}}, \bibinfo {author} {\bibfnamefont {J.}~\bibnamefont {Ghijsen}}, \bibinfo {author} {\bibfnamefont {R.}~\bibnamefont {Sporken}}, \bibinfo {author} {\bibfnamefont {H.}~\bibnamefont {Amara}}, \bibinfo {author} {\bibfnamefont {F.}~\bibnamefont {Ducastelle}},\ and\ \bibinfo {author} {\bibfnamefont
  {J.}~\bibnamefont {Lagoute}},\ }\bibfield  {title} {\bibinfo {title} {Charge transfer and electronic doping in nitrogen-doped graphene},\ }\href {https://doi.org/10.1038/srep14564} {\bibfield  {journal} {\bibinfo  {journal} {Sci. Rep.}\ }\textbf {\bibinfo {volume} {5}},\ \bibinfo {pages} {14564} (\bibinfo {year} {2015})}\BibitemShut {NoStop}%
\bibitem [{\citenamefont {Bostwick}\ \emph {et~al.}(2010)\citenamefont {Bostwick}, \citenamefont {Ohta}, \citenamefont {McChesney}, \citenamefont {Emtsev}, \citenamefont {Speck}, \citenamefont {Seyller}, \citenamefont {Horn}, \citenamefont {Kevan},\ and\ \citenamefont {Rotenberg}}]{bostwick2010interaction}%
  \BibitemOpen
  \bibfield  {author} {\bibinfo {author} {\bibfnamefont {A.}~\bibnamefont {Bostwick}}, \bibinfo {author} {\bibfnamefont {T.}~\bibnamefont {Ohta}}, \bibinfo {author} {\bibfnamefont {J.~L.}\ \bibnamefont {McChesney}}, \bibinfo {author} {\bibfnamefont {K.~V.}\ \bibnamefont {Emtsev}}, \bibinfo {author} {\bibfnamefont {F.}~\bibnamefont {Speck}}, \bibinfo {author} {\bibfnamefont {T.}~\bibnamefont {Seyller}}, \bibinfo {author} {\bibfnamefont {K.}~\bibnamefont {Horn}}, \bibinfo {author} {\bibfnamefont {S.~D.}\ \bibnamefont {Kevan}},\ and\ \bibinfo {author} {\bibfnamefont {E.}~\bibnamefont {Rotenberg}},\ }\bibfield  {title} {\bibinfo {title} {The interaction of quasi-particles in graphene with chemical dopants},\ }\href {https://doi.org/10.1088/1367-2630/12/12/125014} {\bibfield  {journal} {\bibinfo  {journal} {New J. Phys.}\ }\textbf {\bibinfo {volume} {12}},\ \bibinfo {pages} {125014} (\bibinfo {year} {2010})}\BibitemShut {NoStop}%
\bibitem [{\citenamefont {Skrypnyk}\ and\ \citenamefont {Loktev}(2007)}]{skrypnyk2007impurity}%
  \BibitemOpen
  \bibfield  {author} {\bibinfo {author} {\bibfnamefont {Y.~V.}\ \bibnamefont {Skrypnyk}}\ and\ \bibinfo {author} {\bibfnamefont {V.~M.}\ \bibnamefont {Loktev}},\ }\bibfield  {title} {\bibinfo {title} {Impurity induced {Dirac} point smearing in graphene},\ }\href {https://doi.org/10.1063/1.2780170} {\bibfield  {journal} {\bibinfo  {journal} {Low Temp. Phys.}\ }\textbf {\bibinfo {volume} {33}},\ \bibinfo {pages} {762} (\bibinfo {year} {2007})}\BibitemShut {NoStop}%
\bibitem [{\citenamefont {Lin}\ \emph {et~al.}(2021)\citenamefont {Lin}, \citenamefont {Villarreal}, \citenamefont {Achilli}, \citenamefont {Bana}, \citenamefont {Nair}, \citenamefont {Tejeda}, \citenamefont {Verguts}, \citenamefont {De~Gendt}, \citenamefont {Auge}, \citenamefont {Hofs{\"a}ss}, \citenamefont {De~Feyter}, \citenamefont {Di~Santo}, \citenamefont {Petaccia}, \citenamefont {Brems}, \citenamefont {Fratesi},\ and\ \citenamefont {Pereira}}]{lin2021doping}%
  \BibitemOpen
  \bibfield  {author} {\bibinfo {author} {\bibfnamefont {P.-C.}\ \bibnamefont {Lin}}, \bibinfo {author} {\bibfnamefont {R.}~\bibnamefont {Villarreal}}, \bibinfo {author} {\bibfnamefont {S.}~\bibnamefont {Achilli}}, \bibinfo {author} {\bibfnamefont {H.}~\bibnamefont {Bana}}, \bibinfo {author} {\bibfnamefont {M.~N.}\ \bibnamefont {Nair}}, \bibinfo {author} {\bibfnamefont {A.}~\bibnamefont {Tejeda}}, \bibinfo {author} {\bibfnamefont {K.}~\bibnamefont {Verguts}}, \bibinfo {author} {\bibfnamefont {S.}~\bibnamefont {De~Gendt}}, \bibinfo {author} {\bibfnamefont {M.}~\bibnamefont {Auge}}, \bibinfo {author} {\bibfnamefont {H.}~\bibnamefont {Hofs{\"a}ss}}, \bibinfo {author} {\bibfnamefont {S.}~\bibnamefont {De~Feyter}}, \bibinfo {author} {\bibfnamefont {G.}~\bibnamefont {Di~Santo}}, \bibinfo {author} {\bibfnamefont {L.}~\bibnamefont {Petaccia}}, \bibinfo {author} {\bibfnamefont {S.}~\bibnamefont {Brems}}, \bibinfo {author} {\bibfnamefont {G.}~\bibnamefont {Fratesi}},\ and\ \bibinfo {author} {\bibfnamefont {L.~M.~C.}\
  \bibnamefont {Pereira}},\ }\bibfield  {title} {\bibinfo {title} {Doping graphene with substitutional {Mn}},\ }\href {https://doi.org/10.1021/acsnano.1c00139} {\bibfield  {journal} {\bibinfo  {journal} {ACS Nano}\ }\textbf {\bibinfo {volume} {15}},\ \bibinfo {pages} {5449} (\bibinfo {year} {2021})}\BibitemShut {NoStop}%
\bibitem [{\citenamefont {Lin}\ \emph {et~al.}(2022)\citenamefont {Lin}, \citenamefont {Villarreal}, \citenamefont {Bana}, \citenamefont {Zarkua}, \citenamefont {Hendriks}, \citenamefont {Tsai}, \citenamefont {Auge}, \citenamefont {Junge}, \citenamefont {Hofs{\"a}ss}, \citenamefont {Tosi}, \citenamefont {Lacovig}, \citenamefont {Lizzit}, \citenamefont {Zhao}, \citenamefont {Di~Santo}, \citenamefont {Petaccia}, \citenamefont {De~Feyter}, \citenamefont {De~Gendt}, \citenamefont {Brems},\ and\ \citenamefont {Pereira}}]{lin2022thermal}%
  \BibitemOpen
  \bibfield  {author} {\bibinfo {author} {\bibfnamefont {P.-C.}\ \bibnamefont {Lin}}, \bibinfo {author} {\bibfnamefont {R.}~\bibnamefont {Villarreal}}, \bibinfo {author} {\bibfnamefont {H.}~\bibnamefont {Bana}}, \bibinfo {author} {\bibfnamefont {Z.}~\bibnamefont {Zarkua}}, \bibinfo {author} {\bibfnamefont {V.}~\bibnamefont {Hendriks}}, \bibinfo {author} {\bibfnamefont {H.-C.}\ \bibnamefont {Tsai}}, \bibinfo {author} {\bibfnamefont {M.}~\bibnamefont {Auge}}, \bibinfo {author} {\bibfnamefont {F.}~\bibnamefont {Junge}}, \bibinfo {author} {\bibfnamefont {H.}~\bibnamefont {Hofs{\"a}ss}}, \bibinfo {author} {\bibfnamefont {E.}~\bibnamefont {Tosi}}, \bibinfo {author} {\bibfnamefont {P.}~\bibnamefont {Lacovig}}, \bibinfo {author} {\bibfnamefont {S.}~\bibnamefont {Lizzit}}, \bibinfo {author} {\bibfnamefont {W.}~\bibnamefont {Zhao}}, \bibinfo {author} {\bibfnamefont {G.}~\bibnamefont {Di~Santo}}, \bibinfo {author} {\bibfnamefont {L.}~\bibnamefont {Petaccia}}, \bibinfo {author} {\bibfnamefont {S.}~\bibnamefont
  {De~Feyter}}, \bibinfo {author} {\bibfnamefont {S.}~\bibnamefont {De~Gendt}}, \bibinfo {author} {\bibfnamefont {S.}~\bibnamefont {Brems}},\ and\ \bibinfo {author} {\bibfnamefont {L.~M.~C.}\ \bibnamefont {Pereira}},\ }\bibfield  {title} {\bibinfo {title} {Thermal annealing of graphene implanted with {Mn} at ultralow energies: From disordered and contaminated to nearly pristine graphene},\ }\href {https://doi.org/10.1021/acs.jpcc.2c00855} {\bibfield  {journal} {\bibinfo  {journal} {J. Phys. Chem. C}\ }\textbf {\bibinfo {volume} {126}},\ \bibinfo {pages} {10494} (\bibinfo {year} {2022})}\BibitemShut {NoStop}%
\bibitem [{\citenamefont {Villarreal}\ \emph {et~al.}(2024)\citenamefont {Villarreal}, \citenamefont {Zarkua}, \citenamefont {Kretschmer}, \citenamefont {Hendriks}, \citenamefont {Hillen}, \citenamefont {Tsai}, \citenamefont {Junge}, \citenamefont {Nissen}, \citenamefont {Saha}, \citenamefont {Achilli}, \citenamefont {Hofs{\"a}ss}, \citenamefont {Martins}, \citenamefont {De~Ninno}, \citenamefont {Lacovig}, \citenamefont {Lizzit}, \citenamefont {Di~Santo}, \citenamefont {Petaccia}, \citenamefont {De~Feyter}, \citenamefont {De~Gendt}, \citenamefont {Brems}, \citenamefont {Van~de Vondel}, \citenamefont {Krasheninnikov},\ and\ \citenamefont {Pereira}}]{villarreal2024achieving}%
  \BibitemOpen
  \bibfield  {author} {\bibinfo {author} {\bibfnamefont {R.}~\bibnamefont {Villarreal}}, \bibinfo {author} {\bibfnamefont {Z.}~\bibnamefont {Zarkua}}, \bibinfo {author} {\bibfnamefont {S.}~\bibnamefont {Kretschmer}}, \bibinfo {author} {\bibfnamefont {V.}~\bibnamefont {Hendriks}}, \bibinfo {author} {\bibfnamefont {J.}~\bibnamefont {Hillen}}, \bibinfo {author} {\bibfnamefont {H.-C.}\ \bibnamefont {Tsai}}, \bibinfo {author} {\bibfnamefont {F.}~\bibnamefont {Junge}}, \bibinfo {author} {\bibfnamefont {M.}~\bibnamefont {Nissen}}, \bibinfo {author} {\bibfnamefont {T.}~\bibnamefont {Saha}}, \bibinfo {author} {\bibfnamefont {S.}~\bibnamefont {Achilli}}, \bibinfo {author} {\bibfnamefont {H.~C.}\ \bibnamefont {Hofs{\"a}ss}}, \bibinfo {author} {\bibfnamefont {M.}~\bibnamefont {Martins}}, \bibinfo {author} {\bibfnamefont {G.}~\bibnamefont {De~Ninno}}, \bibinfo {author} {\bibfnamefont {P.}~\bibnamefont {Lacovig}}, \bibinfo {author} {\bibfnamefont {S.}~\bibnamefont {Lizzit}}, \bibinfo {author} {\bibfnamefont
  {G.}~\bibnamefont {Di~Santo}}, \bibinfo {author} {\bibfnamefont {L.}~\bibnamefont {Petaccia}}, \bibinfo {author} {\bibfnamefont {S.}~\bibnamefont {De~Feyter}}, \bibinfo {author} {\bibfnamefont {S.}~\bibnamefont {De~Gendt}}, \bibinfo {author} {\bibfnamefont {S.}~\bibnamefont {Brems}}, \bibinfo {author} {\bibfnamefont {J.}~\bibnamefont {Van~de Vondel}}, \bibinfo {author} {\bibfnamefont {A.~V.}\ \bibnamefont {Krasheninnikov}},\ and\ \bibinfo {author} {\bibfnamefont {L.~M.~C.}\ \bibnamefont {Pereira}},\ }\bibfield  {title} {\bibinfo {title} {Achieving high substitutional incorporation in {Mn}-doped graphene},\ }\href {https://doi.org/10.1021/acsnano.4c03475} {\bibfield  {journal} {\bibinfo  {journal} {ACS Nano}\ }\textbf {\bibinfo {volume} {18}},\ \bibinfo {pages} {17815} (\bibinfo {year} {2024})}\BibitemShut {NoStop}%
\bibitem [{\citenamefont {Perdew}\ \emph {et~al.}(1996)\citenamefont {Perdew}, \citenamefont {Burke},\ and\ \citenamefont {Ernzerhof}}]{PBE}%
  \BibitemOpen
  \bibfield  {author} {\bibinfo {author} {\bibfnamefont {J.~P.}\ \bibnamefont {Perdew}}, \bibinfo {author} {\bibfnamefont {K.}~\bibnamefont {Burke}},\ and\ \bibinfo {author} {\bibfnamefont {M.}~\bibnamefont {Ernzerhof}},\ }\bibfield  {title} {\bibinfo {title} {Generalized gradient approximation made simple},\ }\href {https://doi.org/10.1103/PhysRevLett.77.3865} {\bibfield  {journal} {\bibinfo  {journal} {Phys. Rev. Lett.}\ }\textbf {\bibinfo {volume} {77}},\ \bibinfo {pages} {3865} (\bibinfo {year} {1996})}\BibitemShut {NoStop}%
\bibitem [{\citenamefont {Soler}\ \emph {et~al.}(2002)\citenamefont {Soler}, \citenamefont {Artacho}, \citenamefont {Gale}, \citenamefont {Garc{\'i}a}, \citenamefont {Junquera}, \citenamefont {Ordej{\'o}n},\ and\ \citenamefont {S{\'a}nchez-Portal}}]{Sole02}%
  \BibitemOpen
  \bibfield  {author} {\bibinfo {author} {\bibfnamefont {J.~M.}\ \bibnamefont {Soler}}, \bibinfo {author} {\bibfnamefont {E.}~\bibnamefont {Artacho}}, \bibinfo {author} {\bibfnamefont {J.~D.}\ \bibnamefont {Gale}}, \bibinfo {author} {\bibfnamefont {A.}~\bibnamefont {Garc{\'i}a}}, \bibinfo {author} {\bibfnamefont {J.}~\bibnamefont {Junquera}}, \bibinfo {author} {\bibfnamefont {P.}~\bibnamefont {Ordej{\'o}n}},\ and\ \bibinfo {author} {\bibfnamefont {D.}~\bibnamefont {S{\'a}nchez-Portal}},\ }\bibfield  {title} {\bibinfo {title} {The {SIESTA} method for ab initio order-$n$ materials simulation},\ }\href {https://doi.org/10.1088/0953-8984/14/11/302} {\bibfield  {journal} {\bibinfo  {journal} {J. Phys.: Condens. Matter}\ }\textbf {\bibinfo {volume} {14}},\ \bibinfo {pages} {2745} (\bibinfo {year} {2002})}\BibitemShut {NoStop}%
\bibitem [{\citenamefont {Grimme}(2006)}]{Grimme}%
  \BibitemOpen
  \bibfield  {author} {\bibinfo {author} {\bibfnamefont {S.}~\bibnamefont {Grimme}},\ }\bibfield  {title} {\bibinfo {title} {Semiempirical {GGA}-type density functional constructed with a long-range dispersion correction},\ }\href {https://doi.org/10.1002/jcc.20495} {\bibfield  {journal} {\bibinfo  {journal} {J. Comput. Chem.}\ }\textbf {\bibinfo {volume} {27}},\ \bibinfo {pages} {1787} (\bibinfo {year} {2006})}\BibitemShut {NoStop}%
\bibitem [{\citenamefont {Tersoff}\ and\ \citenamefont {Hamann}(1985)}]{Tersoff}%
  \BibitemOpen
  \bibfield  {author} {\bibinfo {author} {\bibfnamefont {J.}~\bibnamefont {Tersoff}}\ and\ \bibinfo {author} {\bibfnamefont {D.~R.}\ \bibnamefont {Hamann}},\ }\bibfield  {title} {\bibinfo {title} {Theory of the scanning tunneling microscope},\ }\href {https://doi.org/10.1103/PhysRevB.31.805} {\bibfield  {journal} {\bibinfo  {journal} {Phys. Rev. B}\ }\textbf {\bibinfo {volume} {31}},\ \bibinfo {pages} {805} (\bibinfo {year} {1985})}\BibitemShut {NoStop}%
\bibitem [{\citenamefont {Shirley}\ \emph {et~al.}(1995)\citenamefont {Shirley}, \citenamefont {Terminello}, \citenamefont {Santoni},\ and\ \citenamefont {Himpsel}}]{shirley1995brillouin}%
  \BibitemOpen
  \bibfield  {author} {\bibinfo {author} {\bibfnamefont {E.~L.}\ \bibnamefont {Shirley}}, \bibinfo {author} {\bibfnamefont {L.~J.}\ \bibnamefont {Terminello}}, \bibinfo {author} {\bibfnamefont {A.}~\bibnamefont {Santoni}},\ and\ \bibinfo {author} {\bibfnamefont {F.~J.}\ \bibnamefont {Himpsel}},\ }\bibfield  {title} {\bibinfo {title} {Brillouin-zone-selection effects in graphite photoelectron angular distributions},\ }\href {https://doi.org/10.1103/PhysRevB.51.13614} {\bibfield  {journal} {\bibinfo  {journal} {Phys. Rev. B}\ }\textbf {\bibinfo {volume} {51}},\ \bibinfo {pages} {13614} (\bibinfo {year} {1995})}\BibitemShut {NoStop}%
\bibitem [{\citenamefont {Mucha-Kruczy{\'n}ski}\ \emph {et~al.}(2008)\citenamefont {Mucha-Kruczy{\'n}ski}, \citenamefont {Tsyplyatyev}, \citenamefont {Grishin}, \citenamefont {McCann}, \citenamefont {Fal'ko}, \citenamefont {Bostwick},\ and\ \citenamefont {Rotenberg}}]{mucha2008characterization}%
  \BibitemOpen
  \bibfield  {author} {\bibinfo {author} {\bibfnamefont {M.}~\bibnamefont {Mucha-Kruczy{\'n}ski}}, \bibinfo {author} {\bibfnamefont {O.}~\bibnamefont {Tsyplyatyev}}, \bibinfo {author} {\bibfnamefont {A.}~\bibnamefont {Grishin}}, \bibinfo {author} {\bibfnamefont {E.}~\bibnamefont {McCann}}, \bibinfo {author} {\bibfnamefont {V.~I.}\ \bibnamefont {Fal'ko}}, \bibinfo {author} {\bibfnamefont {A.}~\bibnamefont {Bostwick}},\ and\ \bibinfo {author} {\bibfnamefont {E.}~\bibnamefont {Rotenberg}},\ }\bibfield  {title} {\bibinfo {title} {Characterization of graphene through anisotropy of constant-energy maps in angle-resolved photoemission},\ }\href {https://doi.org/10.1103/PhysRevB.77.195403} {\bibfield  {journal} {\bibinfo  {journal} {Phys. Rev. B}\ }\textbf {\bibinfo {volume} {77}},\ \bibinfo {pages} {195403} (\bibinfo {year} {2008})}\BibitemShut {NoStop}%
\bibitem [{\citenamefont {Damascelli}\ \emph {et~al.}(2003)\citenamefont {Damascelli}, \citenamefont {Hussain},\ and\ \citenamefont {Shen}}]{damascelli2003angle}%
  \BibitemOpen
  \bibfield  {author} {\bibinfo {author} {\bibfnamefont {A.}~\bibnamefont {Damascelli}}, \bibinfo {author} {\bibfnamefont {Z.}~\bibnamefont {Hussain}},\ and\ \bibinfo {author} {\bibfnamefont {Z.-X.}\ \bibnamefont {Shen}},\ }\bibfield  {title} {\bibinfo {title} {Angle-resolved photoemission studies of the cuprate superconductors},\ }\href {https://doi.org/10.1103/RevModPhys.75.473} {\bibfield  {journal} {\bibinfo  {journal} {Rev. Mod. Phys.}\ }\textbf {\bibinfo {volume} {75}},\ \bibinfo {pages} {473} (\bibinfo {year} {2003})}\BibitemShut {NoStop}%
\bibitem [{\citenamefont {Pramanik}\ \emph {et~al.}(2022)\citenamefont {Pramanik}, \citenamefont {Thakur}, \citenamefont {Singh}, \citenamefont {Willke}, \citenamefont {Wenderoth}, \citenamefont {Hofs{\"a}ss}, \citenamefont {Di~Santo}, \citenamefont {Petaccia},\ and\ \citenamefont {Maiti}}]{Pramanik2022Anomalies}%
  \BibitemOpen
  \bibfield  {author} {\bibinfo {author} {\bibfnamefont {A.}~\bibnamefont {Pramanik}}, \bibinfo {author} {\bibfnamefont {S.}~\bibnamefont {Thakur}}, \bibinfo {author} {\bibfnamefont {B.}~\bibnamefont {Singh}}, \bibinfo {author} {\bibfnamefont {P.}~\bibnamefont {Willke}}, \bibinfo {author} {\bibfnamefont {M.}~\bibnamefont {Wenderoth}}, \bibinfo {author} {\bibfnamefont {H.}~\bibnamefont {Hofs{\"a}ss}}, \bibinfo {author} {\bibfnamefont {G.}~\bibnamefont {Di~Santo}}, \bibinfo {author} {\bibfnamefont {L.}~\bibnamefont {Petaccia}},\ and\ \bibinfo {author} {\bibfnamefont {K.}~\bibnamefont {Maiti}},\ }\bibfield  {title} {\bibinfo {title} {Anomalies at the {Dirac} point in graphene and its hole-doped compositions},\ }\href {https://doi.org/10.1103/PhysRevLett.128.166401} {\bibfield  {journal} {\bibinfo  {journal} {Phys. Rev. Lett.}\ }\textbf {\bibinfo {volume} {128}},\ \bibinfo {pages} {166401} (\bibinfo {year} {2022})}\BibitemShut {NoStop}%
\bibitem [{\citenamefont {Walter}\ \emph {et~al.}(2011{\natexlab{b}})\citenamefont {Walter}, \citenamefont {Nie}, \citenamefont {Bostwick}, \citenamefont {Kim}, \citenamefont {Moreschini}, \citenamefont {Chang}, \citenamefont {Innocenti}, \citenamefont {Horn}, \citenamefont {McCarty},\ and\ \citenamefont {Rotenberg}}]{walter2011copper}%
  \BibitemOpen
  \bibfield  {author} {\bibinfo {author} {\bibfnamefont {A.~L.}\ \bibnamefont {Walter}}, \bibinfo {author} {\bibfnamefont {S.}~\bibnamefont {Nie}}, \bibinfo {author} {\bibfnamefont {A.}~\bibnamefont {Bostwick}}, \bibinfo {author} {\bibfnamefont {K.~S.}\ \bibnamefont {Kim}}, \bibinfo {author} {\bibfnamefont {L.}~\bibnamefont {Moreschini}}, \bibinfo {author} {\bibfnamefont {Y.~J.}\ \bibnamefont {Chang}}, \bibinfo {author} {\bibfnamefont {D.}~\bibnamefont {Innocenti}}, \bibinfo {author} {\bibfnamefont {K.}~\bibnamefont {Horn}}, \bibinfo {author} {\bibfnamefont {K.~F.}\ \bibnamefont {McCarty}},\ and\ \bibinfo {author} {\bibfnamefont {E.}~\bibnamefont {Rotenberg}},\ }\bibfield  {title} {\bibinfo {title} {Electronic structure of graphene on single-crystal copper substrates},\ }\href {https://doi.org/10.1103/PhysRevB.84.195443} {\bibfield  {journal} {\bibinfo  {journal} {Phys. Rev. B}\ }\textbf {\bibinfo {volume} {84}},\ \bibinfo {pages} {195443} (\bibinfo {year} {2011}{\natexlab{b}})}\BibitemShut {NoStop}%
\bibitem [{\citenamefont {Katsnelson}\ and\ \citenamefont {Geim}(2008)}]{katsnelson2008electron}%
  \BibitemOpen
  \bibfield  {author} {\bibinfo {author} {\bibfnamefont {M.~I.}\ \bibnamefont {Katsnelson}}\ and\ \bibinfo {author} {\bibfnamefont {A.~K.}\ \bibnamefont {Geim}},\ }\bibfield  {title} {\bibinfo {title} {Electron scattering on microscopic corrugations in graphene},\ }\href {https://doi.org/10.1098/rsta.2007.2157} {\bibfield  {journal} {\bibinfo  {journal} {Philos. Trans. R. Soc. A}\ }\textbf {\bibinfo {volume} {366}},\ \bibinfo {pages} {195} (\bibinfo {year} {2008})}\BibitemShut {NoStop}%
\bibitem [{\citenamefont {Zarkua}\ \emph {et~al.}(2026)\citenamefont {Zarkua}, \citenamefont {Smeyers}, \citenamefont {Seliverstov}, \citenamefont {Villarreal}, \citenamefont {Lotfy}, \citenamefont {Joris}, \citenamefont {Saad}, \citenamefont {Tsai}, \citenamefont {{De Gendt}}, \citenamefont {Brems}, \citenamefont {{De Feyter}}, \citenamefont {Junge}, \citenamefont {Hofs{\"a}ss}, \citenamefont {{Di Santo}}, \citenamefont {Petaccia}, \citenamefont {Achilli}, \citenamefont {{\AA}hlgren}, \citenamefont {Peeters}, \citenamefont {Milo{\v{s}}evi{\'c}}, \citenamefont {Covaci},\ and\ \citenamefont {Pereira}}]{zarkua2026electronic}%
  \BibitemOpen
  \bibfield  {author} {\bibinfo {author} {\bibfnamefont {Z.}~\bibnamefont {Zarkua}}, \bibinfo {author} {\bibfnamefont {R.}~\bibnamefont {Smeyers}}, \bibinfo {author} {\bibfnamefont {A.}~\bibnamefont {Seliverstov}}, \bibinfo {author} {\bibfnamefont {R.}~\bibnamefont {Villarreal}}, \bibinfo {author} {\bibfnamefont {A.~S.}\ \bibnamefont {Lotfy}}, \bibinfo {author} {\bibfnamefont {R.}~\bibnamefont {Joris}}, \bibinfo {author} {\bibfnamefont {M.}~\bibnamefont {Saad}}, \bibinfo {author} {\bibfnamefont {H.-C.}\ \bibnamefont {Tsai}}, \bibinfo {author} {\bibfnamefont {S.}~\bibnamefont {{De Gendt}}}, \bibinfo {author} {\bibfnamefont {S.}~\bibnamefont {Brems}}, \bibinfo {author} {\bibfnamefont {S.}~\bibnamefont {{De Feyter}}}, \bibinfo {author} {\bibfnamefont {F.}~\bibnamefont {Junge}}, \bibinfo {author} {\bibfnamefont {H.}~\bibnamefont {Hofs{\"a}ss}}, \bibinfo {author} {\bibfnamefont {G.}~\bibnamefont {{Di Santo}}}, \bibinfo {author} {\bibfnamefont {L.}~\bibnamefont {Petaccia}}, \bibinfo {author} {\bibfnamefont
  {S.}~\bibnamefont {Achilli}}, \bibinfo {author} {\bibfnamefont {E.~H.}\ \bibnamefont {{\AA}hlgren}}, \bibinfo {author} {\bibfnamefont {F.~M.}\ \bibnamefont {Peeters}}, \bibinfo {author} {\bibfnamefont {M.~V.}\ \bibnamefont {Milo{\v{s}}evi{\'c}}}, \bibinfo {author} {\bibfnamefont {L.}~\bibnamefont {Covaci}},\ and\ \bibinfo {author} {\bibfnamefont {L.~M.~C.}\ \bibnamefont {Pereira}},\ }\bibfield  {title} {\bibinfo {title} {Electronic effects of localized strain in graphene},\ }\href {https://doi.org/10.1016/j.carbon.2026.121343} {\bibfield  {journal} {\bibinfo  {journal} {Carbon}\ }\textbf {\bibinfo {volume} {251}},\ \bibinfo {pages} {121343} (\bibinfo {year} {2026})}\BibitemShut {NoStop}%
\end{thebibliography}%
\end{document}